\documentclass[11pt,a4paper]{article}
\usepackage[T1]{fontenc}
\usepackage{lmodern}
\usepackage{jheppub}
\makeatletter
\let\@fpheader\@empty
\makeatother
\usepackage{amsthm,mathtools,booktabs,array}
\usepackage{tikz}
\usetikzlibrary{arrows.meta,calc,decorations.markings,decorations.pathmorphing,positioning}

\definecolor{ampSlate}{RGB}{38,50,66}
\definecolor{ampBlue}{RGB}{32,96,158}
\definecolor{ampTeal}{RGB}{24,126,122}
\definecolor{ampRed}{RGB}{184,46,56}
\definecolor{ampMuted}{RGB}{105,115,128}
\definecolor{ampSoftFill}{RGB}{236,243,251}
\definecolor{ampBlobFill}{RGB}{240,243,247}
\definecolor{ampDiskFill}{RGB}{247,249,252}

\tikzset{
  hard line/.style={draw=ampSlate, line width=1.05pt},
  soft line/.style={draw=ampBlue, line width=1.05pt, densely dashed},
  neg leg/.style={draw=ampRed, line width=1.15pt},
  pos leg/.style={draw=ampSlate, line width=0.85pt},
  soft grav/.style={draw=ampTeal, line width=0.95pt,
    decorate, decoration={snake, amplitude=1.4pt, segment length=5.5pt}},
  cut line/.style={draw=ampRed, line width=1.05pt, dashed},
  hard blob/.style={circle, draw=ampSlate, fill=ampBlobFill,
    line width=0.95pt, inner sep=0pt},
  soft blob/.style={circle, draw=ampBlue, fill=ampSoftFill,
    line width=0.95pt, densely dashed, inner sep=0pt},
  star node/.style={circle, draw=ampRed, fill=ampRed, inner sep=1.7pt},
  mid arrow/.style={postaction={decorate},
    decoration={markings, mark=at position 0.34 with {\arrow{Stealth[length=4.5pt,width=3.5pt]}}}},
}

\hypersetup{pdftitle={MHV graviton amplitudes at and near tree zeros in maximal supergravity},pdfauthor={Zhen Bi and Yaobo Zhang},pdfkeywords={Scattering Amplitudes, Supergravity Models, Extended Supersymmetry}}
\newtheorem{theorem}{Theorem}

\numberwithin{equation}{section}
\allowdisplaybreaks[1]
\title{MHV graviton amplitudes at and near tree zeros in maximal supergravity}
\author[a]{Zhen Bi}
\author[b]{and Yaobo Zhang}
\affiliation[a]{College of Information Engineering, Huzhou Normal University,\\
759 Erhuandong Road, Huzhou, Zhejiang 313000, China}
\affiliation[b]{School of Physics, Ningxia University, Yinchuan 750021, China}
\emailAdd{yaobozhang@nxu.edu.cn}
\abstract{We study loop corrections to tree-level MHV graviton zeros in
four-dimensional maximal supergravity. Near a split-signature zero where
the positive-helicity square spinors align, individual box contributions
can appear below the generic tree power at six or more points. We prove
that these terms cancel in the full one-loop finite part. At the first
allowed power the double logarithm cancels, and the single-logarithm
coefficient is the leading tree coefficient times a purely imaginary
factor fixed by the hard invariants. In real Lorentzian $2\to n-2$
scattering with $n\ge5$, every regular MHV tree zero is planar. There the
one-loop amplitude is infrared finite and nonzero for every choice of the
negative-helicity pair. Other energy assignments admit symmetry-enforced
common zeros. We also construct split-signature paths on which the tree
vanishes arbitrarily many powers faster than the one-loop finite part.
For each such power gap we compare two paths at the same subtraction
scale and hard invariant. On at least one of them the two-loop finite
part starts at or below the one-loop power.}
\begin{document}
\maketitle

\section{Introduction}

How fast does a one-loop MHV graviton amplitude vanish as the external
momenta approach a helicity zero? Take six gravitons in split signature
with negative-helicity labels $1,a$. Along the path let $[ij]=O(t)$
for every positive-helicity pair and
$\langle1a\rangle=\delta_1 t+O(t^2)$ with $\delta_1\ne0$. Strip the common
external factor $\langle1a\rangle^8$ from each term in the
Bern--Dixon--Perelstein--Rozowsky box expansion~\cite{BDDPR}. A single
box contribution can then have a nonzero constant term, while the stripped
tree vanishes linearly on a generic path. The sum over all $105$ boxes
cancels the constant exactly. The first term on a generic path is
$t\log t$, or $t^9\log t$ with the external helicity factor restored.
At $n$ points the first allowed stripped power is $n-5$. Six is the
first multiplicity at which even a constant must cancel. At ten points
the cancellation extends through $1,t,t^2,t^3,t^4$.
Figure~\ref{fig:one-loop-boxes} shows which boxes can contribute at
each order. Section~\ref{one:examples} checks the cancellation by direct
summation through ten points.

In real $2\to n-2$ scattering we can set the tree exactly to zero with
all null directions distinct and all physical channels off their poles.
Does the integrated one-loop amplitude vanish there? For $n\ge5$ we show
that every such MHV tree zero is planar. The one-loop infrared pole
vanishes with the tree, leaving a finite, nonzero amplitude for every
choice of the negative-helicity pair. Its absorptive part is strictly
negative in the real planar spin frame of section~\ref{sec:physical}.
In the five-point example with both negative helicities on the incoming
beams, it equals $-81\pi\log 2$.

Throughout the paper we work with the dimensionally regulated, integrated
amplitude in $D=4-2\epsilon$ with four-dimensional external helicities.
We strip the momentum-conserving delta function, the gravitational
coupling, and common loop factors according to the normalization
in~\eqref{one:boxes}, and write
\[
  \mathcal T=M^{(0)},\qquad F_L=[\epsilon^0]M^{(L)},\qquad q=n+3.
\]
We call $F_L$ the finite part. On a path we take the Laurent coefficient
at fixed $t>0$, then let $t\downarrow0$. At two loops this coefficient
includes finite terms generated by lower-loop infrared poles. All momenta are outgoing,
$p_i=\lambda_i\widetilde\lambda_i^{\mathsf T}$ with $\sum_i p_i=0$,
$s_I=\det\sum_{i\in I}p_i$, and $s_{ij}=\langle ij\rangle[ij]$, where the
brackets denote ordered two-spinor determinants.

\paragraph{Approaching a split-signature helicity zero.}
The helicity-zero rule of Jones and Paranjape allows either all
positive-helicity square spinors or all negative-helicity angle spinors
to align~\cite{JP}. Choosing the former branch, write
$\widetilde\lambda_i(0)=c_i\widetilde\eta$ for $i\in P_+$. At $t=0$
momentum conservation reads
\[
 -(p_1+p_a)=\sum_{i\in P_+}p_i
     =\Big(\sum_{i\in P_+}c_i\lambda_i\Big)\widetilde\eta^{\mathsf T},
 \qquad s_{1a}=\langle1a\rangle[1a]=0 .
\]
Because the right-hand side has rank at most one, requiring
$[1a](0)\ne0$ forces $\langle1a\rangle(0)=0$ as well. We assume distinct
positive angle directions, nonvanishing mixed brackets, and nonzero
endpoint invariants for every nontrivial partition separating $1$ and $a$.
Here the momenta are split-real; for real Lorentzian momenta the two
chiral brackets are complex conjugates up to energy signs and phases, so
they vanish together, whereas the regular Lorentzian zeros of
section~\ref{sec:physical} have distinct directions for all $n$ legs.

The generic tree scaling follows directly from the Hodges determinant and
its spanning-tree expansion~\cite{Hodges,FH}: the MHV tree carries an
overall $\langle1a\rangle^6$ together with $n-3$ positive-pair edges
$[ij]=O(t)$, yielding $t^q$ with the weighted cofactor $T_0$
in~\eqref{one:cofactor}. On any regular transverse path we find
\begin{align*}
 \mathcal T &=t^q\bigl(T_0+O(t)\bigr),&
 F_1 &=t^q\bigl[-\sigma_0 T_0\log t+B+o(1)\bigr],\\[-2pt]
 && \sigma_0 &=-i\pi\sum_{j\in P_+}|s_{1j}(0)|.
\end{align*}
All lower powers and the order-$q$ double logarithm cancel in the full box
sum. Whenever $T_0\ne0$, the loop-to-tree ratio grows logarithmically,
$F_1/\mathcal T=-\sigma_0\log t+O(1)$; on the six-point exponential path
of section~\ref{one:examples} it equals $(16i\pi/7)\log t+O(1)$.

The absolute values in $\sigma_0$ come directly from the Feynman
prescription $L(s)=\log((-s-i0)/\mu^2)$. Since each hard pair obeys
$b_j=s_{1j}(0)=-s_{aj}(0)$, the opposite signs cancel the real logarithms
and add the imaginary parts coherently,
$b_j[L(-b_j)-L(b_j)]=i\pi|b_j|$. Because all positive-pair invariants
vanish at $t=0$, any channel separating $1$ and $a$ has endpoint value
$s_{\{1\}\cup S}(0)=\sum_{j\in S}b_j$. Together with $\sum_j b_j=0$, this
means the subset $J=\{j\in P_+:b_j>0\}$ uniquely maximizes this invariant,
giving $2s_{\{1\}\cup J}(0)=\sum_j|b_j|$. Thus a single extremal channel
packages all hard-pair phases within a fixed sign region.

\paragraph{Relation to hidden zeros and cuts.}
An ordered four-pion amplitude gives a simple example of a tree zero
away from planar poles~\cite{AHCDFH}. In the massless nonlinear sigma
model the tree is proportional to $-u=-s_{13}$ and vanishes at $u=0$
with $s=s_{12}=-s_{23}>0$. On the $s$-channel cut, legs $1$ and $3$
belong to different trees. Their product has a nonzero angular average,
as illustrated in the ancillary figure. For the $(1234)$ box color
coefficient with the overall $i$ removed, the imaginary part at the
zero is $s^2/(96\pi f_\pi^4)$, so the magnitude of the loop-to-tree
ratio grows as $1/|u|$ at fixed $s$~\cite[Eq.~(4.16)]{CarrascoPavao}.

In ordered theories, hidden zeros have also been used to constrain
loop-level integrands~\cite{BR,Rodina26}. Here we work after loop
integration in gravity, where all factorization channels participate
simultaneously~\cite{HYZ} and the MHV tree divisor is governed by the
Hodges determinant~\cite{KOPTZ,MZ,KVY1,KVY2}. On the two-incoming
Lorentzian slice of that divisor, section~\ref{sec:physical} connects the
known planar tree zeros~\cite{Planar} to the definite sign of the
integrated absorptive part. Away from $2\to n-2$ kinematics, a seven-point
$\mathbb Z_3$-symmetric configuration enforces a common zero
$\mathcal T=F_1=0$, while deformations preserving both the energy
signature and the excess Hodges corank move $F_1$ away from zero in two
independent real directions.

\paragraph{A delayed tree.}
What happens when the leading cofactor $T_0$ vanishes? The
$t^q\log t$ term drops out of $F_1$, but the constant $B$ at the same
power can survive. Section~\ref{sec:overtaking} gives a six-point path
on which the tree starts at $t^{11}$ and the one-loop finite part at
$t^9$. The transverse scale is still fixed by
$\langle1a\rangle=\delta_1 t+O(t^2)$. The two extra powers come from
the choice of direction.

More generally, for each $n\ge6$ and integer $H\ge1$ we construct two
paths that follow a simple tree-zero branch through order $H-1$. Both
are strictly positive for small $t>0$, meaning that all ordered
four-minors of the real momentum twistors are positive in the chosen
frame and auxiliary cyclic order. This places the tuned paths in the
interior of the corresponding positive twistor region. On both paths
$\operatorname{ord}_t\mathcal T=q+H$ and $\operatorname{ord}_t F_1=q$.
Here $\operatorname{ord}_t$ counts the first power of $t$ with a nonzero
coefficient polynomial in $\log t$ in a regular analytic spin frame.
We compare the two paths at fixed subtraction scale and a common
prescribed mixed invariant $S>0$, using the infrared poles of
section~\ref{sec:ir} and the response of $F_2$ to a momentum dilation. On at least one of them $\operatorname{ord}_t F_2\le q$.
To examine one two-loop channel directly, we turn to the negative-helicity
pair cut in section~\ref{sec:cut}. There we evaluate its four-point
one-loop contribution and relate the scale derivative of that integral
to the pole terms.

\section{One-loop cancellations near a split-signature helicity zero}
\label{sec:one}

\subsection{A tree cofactor and the normalized boxes}

The regular transverse limit used here has $N=n-2$ positive legs,
$q=n+3$, and real analytic conserving spinors. At $t=0$ the negative
angle spinors are parallel, $\langle1a\rangle=\delta_1 t+O(t^2)$ with
$\delta_1\ne0$, and all positive square spinors are parallel, so
$[ij]=O(t)$ for $i,j\in P_+$. We require $[1a](0)\ne0$, distinct
positive angle directions, and nonvanishing mixed angle and square
brackets at the endpoint. We call a channel hard if its invariant has
a nonzero endpoint value and soft if that value vanishes. Every
partition separating $1$ and $a$ with at least two legs on each side
must be hard. Soft channels may vanish at different rates.

For a nonempty set $M$, fix reference legs $b,c$ and set
$w_i=\langle bi\rangle\langle ic\rangle$ and
$g_{ij}=[ij]/\langle ij\rangle$. The half-soft function is the
weighted spanning-tree sum~\cite{BDDPR,FH},
\begin{equation}
 h(b,M,c)=\frac{1}{\prod_{i\in M}w_i^2}
   \sum_{\text{trees }G\text{ on }M}
       \prod_{ij\in G}w_iw_jg_{ij} .
 \label{one:halfsoft}
\end{equation}
A singleton has one empty tree. In this normalization the MHV component
is $\mathcal T=\langle1a\rangle^6 h(1,P_+,a)$~\cite{Hodges,FH}.
Let $v_i=\langle1i\rangle\langle ai\rangle|_0$ and form the
Laplacian $L_*$ with off-diagonals $-v_iv_j\partial_t[ij]|_0/
\langle ij\rangle_0$ and vanishing row sums. All $N-1$ edges of a
positive spanning tree carry at least one power of $t$. The matrix-tree
theorem gives
\begin{equation}
 \mathcal T=t^q(T_0+O(t)),\qquad
 T_0=\delta_1^{\,6}\frac{\det\widehat L_*}{\prod_{i\in P_+}v_i^2},
 \label{one:cofactor}
\end{equation}
where the hat deletes any one row and the same column. The cofactor can
vanish even when every edge has a nonzero linear coefficient.

At any regular endpoint one can also see directly that $T_0$ is
generically nonzero among conserving transverse derivatives. Normalize
the two negative angles to $(1,0)$ at zero, write the other angles as
$(x_i,1)$ and the positive square spinors there as $q_i\eta$. Set
$\lambda_a=(1,t)$, $\widetilde\lambda_a=\nu$ and
$\widetilde\lambda_i=q_i\eta+tp_i\nu$, where $\nu$ is the endpoint
square spinor of $a$ and $\sum_i p_i=-1$. The second angle row of
conservation holds since $\sum q_i=0$. The first fixes
$\widetilde\lambda_1=-\nu-\sum_i x_i\widetilde\lambda_i$.
Choosing just one nonzero $p_h=-1$ makes the star through $h$ the
only surviving tree in \eqref{one:cofactor}. Thus the cofactor is a
nontrivial polynomial in the first derivative. This argument concerns
unrestricted conserving directions. Strictly positive examples are
given in section~\ref{one:examples}.

Take each unordered pair of massless corners $b,c$ and each unordered
split of the remaining labels into two nonempty corners $M,M'$ once.
Write
\begin{align}
 S&=s_{M\cup\{b\}},& T'&=s_{M\cup\{c\}},&P&=s_M,& Q&=s_{M'},\nonumber\\
 D&=ST'-PQ,&u&=-s_{bc}/D,&(\sigma_S,\sigma_{T'},\sigma_P,\sigma_Q)&=(1,1,-1,-1).
 \label{one:boxkin}
\end{align}
Here $s_{ij}=\langle ij\rangle[ij]$ and the identity
$S+T'-P-Q=-s_{bc}$ follows from conservation. The reference scale is
$\mu^2=1$. We strip the coupling and the common one-loop prefactor,
including $r_\Gamma$, and normalize the one-loop finite part
$F_1=[\epsilon^0]M^{(1)}$ by
\begin{align}
 F_1 &=\sum_{\text{boxes }\gamma} C_\gamma f_\gamma,\qquad
 C_\gamma=\langle1a\rangle^8D\,h(b,M,c)h(c,M',b),\label{one:boxes}\\
 f_\gamma&=-\sum_{X=S,T',P,Q}\sigma_X
    \left[\tfrac12 L_X^2+\operatorname{Li}_2(1-u(X+i0))\right],
 \qquad L_X=\log(-X-i0).\label{one:kernel}
\end{align}
All four positions count with multiplicity. Equation~\eqref{one:kernel}
uses the Feynman-sheet boundary values of the one-mass and two-mass-easy
scalar integrals~\cite{DN}. For a massless corner the regulated integral
omits the scale-free power $(-X)^{-\epsilon}$, so the $L_X^2$ term is
absent there and the dilogarithm contributes $\operatorname{Li}_2(1)$.

\paragraph{The box coefficient on the quadruple cut.}
The four-dimensional box reduction through $\epsilon^0$~\cite{BV}
leaves the identification of the rational coefficients. Bern et
al.~\cite[Eq.~(6.18)]{BDDPR} inferred their all-multiplicity form
from low-point cuts and factorization. We can recover it directly
from the MHV quadruple cut~\cite[Sec.~3]{KST} and the Hodges
tree~\cite{Hodges,FH}. Work at generic kinematics before taking the
helicity-zero limit. The external MHV superamplitude has Grassmann
degree 16, and the four internal state sums remove degree 32. At least
two corners must therefore be three-point anti-MHV trees of degree
eight. Two adjacent such corners would align the angle spinors of
their external legs. Thus the anti-MHV corners are opposite, at $b,c$,
and the remaining two corners are MHV.

On this solution the angle spinors of the cut lines beside $b$ can
both be chosen as $|b\rangle$, and likewise at $c$. With
$K=\sum_{i\in M}p_i$, the two cut momenta outgoing from corner $M$
then have square spinors $[\rho|,[\tau|$ fixed by
\begin{equation}
 |b\rangle[\rho|+|c\rangle[\tau|=-K,\qquad
 [\rho|=-\frac{\langle c|K}{\langle cb\rangle},\qquad
 [\tau|=-\frac{\langle b|K}{\langle bc\rangle}.
 \label{one:cutspinors}
\end{equation}
The cut momenta outgoing from $M'$ have the same two angle spinors
and square spinors $[b|-[\rho|$ and $[c|-[\tau|$. To evaluate a
corner tree with the external set $M$, strip its supermomentum
delta function $\delta^{16}(\mathcal Q)$. In the Hodges reduced determinant,
delete the two cut legs and a third leg $r\in M$ and use the cut legs
as reference spinors. Multiplying the remaining rows and columns by
$w_i$ gives the minor of the tree Laplacian $L_M$ with edge weights
$w_iw_j g_{ij}$. The deleted angle-bracket product squared is
$\langle bc\rangle^2 w_r^2$. Hence the stripped tree is
\begin{equation}
 \frac{\det (L_M)_{\widehat r,\widehat r}}
 {\langle bc\rangle^2\prod_{i\in M}w_i^2}
 =\frac{h(b,M,c)}{\langle bc\rangle^2}.
 \label{one:cutcorner}
\end{equation}
For $M=\{i\}$ the minor is the empty determinant. Equation~\eqref{one:cutspinors}
then aligns both cut square spinors with $[i|$, and
\eqref{one:cutcorner} is precisely the stripped MHV three-point tree
$1/(\langle bc\rangle^2\langle ci\rangle^2\langle ib\rangle^2)$.

To perform the state sum, extract the external component $b^-,c^-$.
Each negative graviton uses all eight Grassmann variables of its
anti-MHV corner, so only positive gravitons occur on its two internal
legs. Both cut legs of each MHV tree consequently have negative
helicity. Its component is $\langle bc\rangle^8$ times
\eqref{one:cutcorner}. At $b$, the three square brackets in the
anti-MHV graviton tree agree up to signs by conservation, giving
$[b\rho]^2$. The other anti-MHV tree gives $[c\tau]^2$. The
two-dimensional determinant identity
$\langle b|K|c]\langle c|K|b]=ST'-PQ$ gives
$[b\rho]^2[c\tau]^2=D^2/\langle bc\rangle^4$.
The four trees thus give
$\langle bc\rangle^8D^2 h(b,M,c)h(c,M',b)$.
The external MHV Ward identity replaces $\langle bc\rangle^8$
by $\langle1a\rangle^8$ for the component of interest. The scalar
box supplies $1/D$ on conversion to the dimensionless kernel
\eqref{one:kernel}. The common factors, including the $1/2$ in the
quadruple-cut average with a single contributing solution, are absorbed
in our normalization. This gives \eqref{one:boxes} both for the
five-point one-mass boxes and at all higher multiplicities.

\subsection{Why every power below \texorpdfstring{$n+3$}{n+3} cancels}

Call a box a \emph{star box} if $b,c\in P_+$ and its two other
corners contain one negative leg each. In a corner $\{v\}\cup A$,
only the tree with centre $v$ survives at the endpoint. Its value is
$w_v^{|A|-2}\prod_{i\in A}g_{vi}/w_i$. The Gram factor $D$ is a unit
there, so $C_\gamma$ starts at $t^8$. After removing
$\langle1a\rangle^8$, its coefficient is of order one while the tree
starts generically at $t^{n-5}$. Six points is the first multiplicity
at which an individual star coefficient precedes the tree. The other boxes have the
following bounds.

\begin{itemize}
\item With one negative massless corner, write $p$ for the number of
positive companions of the other negative leg, and $r$ for the size of
the all-positive corner. Then $p+r=n-3$. Each edge in the first tree
costs a power of $t$, and its negative vertex contributes $w_v^{-2}=
O(t^{-2})$. The half-soft factors are $O(t^{p-2})$ and $O(t^{r-1})$.
One diagonal channel and the all-positive corner channel are soft,
so $D=O(t)$ and $C_\gamma=O(t^q)$.
\item With both negatives massless the trees contain $n-4$ positive
edges in total, hence $C_\gamma=O(t^{q+1})$.
\item With both negatives in one corner let $p$ be the number of their
positive companions and $r=n-4-p$ the opposite corner size. All four
channels are $O(t)$. Conservation and $\lambda_a(0)\parallel\lambda_1(0)$
imply $w_1w_jg_{1j}+w_aw_jg_{aj}=O(t)$. After deletion of vertex $1$
the $p\times p$ positive block of the tree Laplacian is $O(t)$, the
mixed entries are bounded and its remaining diagonal is $O(t^{-1})$.
The determinant is $O(t^{p-1})$. The opposite tree contributes
$O(t^{r-1})$ and $D=O(t^2)$, so again $C_\gamma=O(t^{q+1})$.
\end{itemize}

\begin{figure}[htbp]
\centering
\resizebox{\textwidth}{!}{%
\begin{tikzpicture}[>=Stealth, font=\small]

\begin{scope}[shift={(0,0)}]
  \node[font=\bfseries, text=ampSlate] at (0.90, 3.25) {(a) Star box};

  \draw[hard line] (0,0) rectangle (1.80,1.80);

  \draw[pos leg] (0,1.80) -- (-0.62,2.30);
  \node[text=ampSlate, anchor=south east, inner sep=2pt] at (-0.60,2.28) {$b^+$};
  \draw[pos leg] (1.80,0) -- (2.42,-0.50);
  \node[text=ampSlate, anchor=north west, inner sep=2pt] at (2.40,-0.48) {$c^+$};

  \draw[neg leg] (1.80,1.80) -- (2.25,2.25);
  \node[star node] (s1) at (2.25,2.25) {};
  \node[text=ampRed, font=\footnotesize, anchor=south east, inner sep=2pt] at (2.22,2.28) {$1^-$};
  \draw[pos leg] (s1) -- (2.68,2.78);
  \draw[pos leg] (s1) -- (2.94,2.40);
  \draw[pos leg] (s1) -- (2.84,1.94);
  \node[text=ampSlate, font=\footnotesize, anchor=west] at (2.94,2.38) {$A$};

  \draw[neg leg] (0,0) -- (-0.45,-0.45);
  \node[star node] (sa) at (-0.45,-0.45) {};
  \node[text=ampRed, font=\footnotesize, anchor=north west, inner sep=2pt] at (-0.42,-0.48) {$a^-$};
  \draw[pos leg] (sa) -- (-0.98,-0.16);
  \draw[pos leg] (sa) -- (-1.12,-0.58);
  \draw[pos leg] (sa) -- (-0.84,-0.98);
  \node[text=ampSlate, font=\footnotesize, anchor=east] at (-1.12,-0.58) {$A'$};

  \node[text=ampSlate, font=\footnotesize, align=center] at (0.90,0.90) {analytic\\kernel};

  \node[text=ampSlate, font=\footnotesize] at (0.90,-1.52) {$D(0)\ne0,\quad \operatorname{ord}_t C_\gamma=8$};
  \node[text=ampMuted, font=\footnotesize] at (0.90,-1.98) {star sum $=O(t^q)$};
\end{scope}

\begin{scope}[shift={(5.95,0)}]
  \node[font=\bfseries, text=ampSlate] at (0.90, 3.25) {(b) Mixed helicities at $b,c$};

  \draw[hard line] (0,0) rectangle (1.80,1.80);

  \draw[neg leg] (0,1.80) -- (-0.62,2.30);
  \node[text=ampRed, anchor=south east, inner sep=2pt] at (-0.60,2.28) {$b^-$};
  \draw[pos leg] (1.80,0) -- (2.42,-0.50);
  \node[text=ampSlate, anchor=north west, inner sep=2pt] at (2.40,-0.48) {$c^+$};

  \draw[neg leg] (1.80,1.80) -- (2.48,2.62);
  \draw[pos leg] (1.80,1.80) -- (2.78,2.28);
  \draw[pos leg] (1.80,1.80) -- (2.74,1.86);
  \node[hard blob, minimum size=15pt] at (1.80,1.80) {};
  \node[text=ampSlate, font=\footnotesize, anchor=west] at (2.78,2.26) {$\{v^-\}\!\cup\!M_+$};
  \node[text=ampSlate, font=\footnotesize, anchor=north west] at (2.06,1.60) {$P$ hard};

  \draw[pos leg] (0,0) -- (-0.82,-0.16);
  \draw[pos leg] (0,0) -- (-0.86,-0.58);
  \draw[pos leg] (0,0) -- (-0.54,-0.92);
  \node[soft blob, minimum size=15pt] at (0,0) {};
  \node[text=ampBlue, font=\footnotesize, anchor=east] at (-0.88,-0.56) {$M'_+$};
  \node[text=ampBlue, font=\footnotesize, anchor=south east] at (-0.23,0.14) {$Q$ soft};

  \node[text=ampSlate, font=\footnotesize] at (0.90,1.09) {$T'$ hard};
  \node[text=ampBlue, font=\footnotesize] at (0.90,0.71) {$S$ soft};

  \node[text=ampSlate, font=\footnotesize] at (0.90,-1.52) {$D=O(t),\quad C_\gamma=O(t^q)$};
  \node[text=ampBlue, font=\footnotesize] at (0.90,-1.98) {source of $t^q\log t$};
\end{scope}

\begin{scope}[shift={(12.15,0)}]
  \node[font=\bfseries, text=ampSlate] at (0.90, 3.25) {(c) Suppressed boxes};

  \draw[hard line] (0,0) rectangle (1.80,1.80);

  \draw[pos leg] (0,1.80) -- (-0.62,2.30);
  \node[text=ampSlate, anchor=south east, inner sep=2pt] at (-0.60,2.28) {$b^+$};
  \draw[pos leg] (1.80,0) -- (2.42,-0.50);
  \node[text=ampSlate, anchor=north west, inner sep=2pt] at (2.40,-0.48) {$c^+$};

  \draw[neg leg] (1.80,1.80) -- (2.34,2.68);
  \draw[neg leg] (1.80,1.80) -- (2.68,2.40);
  \draw[pos leg] (1.80,1.80) -- (2.78,1.98);
  \node[soft blob, minimum size=15pt] at (1.80,1.80) {};
  \node[text=ampSlate, font=\footnotesize, anchor=west] at (2.74,2.34) {$\{1^-,a^-\}\!\cup\!M_+$};

  \draw[pos leg] (0,0) -- (-0.82,-0.16);
  \draw[pos leg] (0,0) -- (-0.86,-0.58);
  \draw[pos leg] (0,0) -- (-0.54,-0.92);
  \node[soft blob, minimum size=15pt] at (0,0) {};
  \node[text=ampBlue, font=\footnotesize, anchor=east] at (-0.88,-0.56) {$M'_+$};

  \node[text=ampBlue, font=\footnotesize, align=center] at (0.90,0.90) {four soft\\channels};

  \node[text=ampSlate, font=\footnotesize] at (0.90,-1.52) {$D=O(t^2),\quad C_\gamma=O(t^{q+1})$};
\end{scope}

\path ([xshift=-4pt,yshift=-4pt]current bounding box.south west)
      ([xshift=4pt,yshift=4pt]current bounding box.north east);

\end{tikzpicture}%
}
\caption{One-loop box coefficients in~\eqref{one:boxes} for $q=n+3$.
\textbf{(a)}~The two half-soft trees at $t=0$ are stars centred on the
negative legs. Their sum over boxes cancels at every power below $q$
by Theorem~\ref{one:low}.
\textbf{(b)}~Only boxes with one negative and one positive massless
corner can supply the logarithm in Theorem~\ref{one:branch}. Summing
their hard-pair phases selects the channel with side $\{1\}\cup J$
in~\eqref{one:selected}. The diagram labels the external invariants
of~\eqref{one:boxkin} for a nonsingleton corner containing $v^-$.
For $M_+=\varnothing$, $P$ vanishes identically.
\textbf{(c)}~Both negatives lie in one corner. Putting them instead at
the two massless corners also gives $C_\gamma=O(t^{q+1})$, through the
$n-4$ positive--positive tree edges.}
\label{fig:one-loop-boxes}
\end{figure}
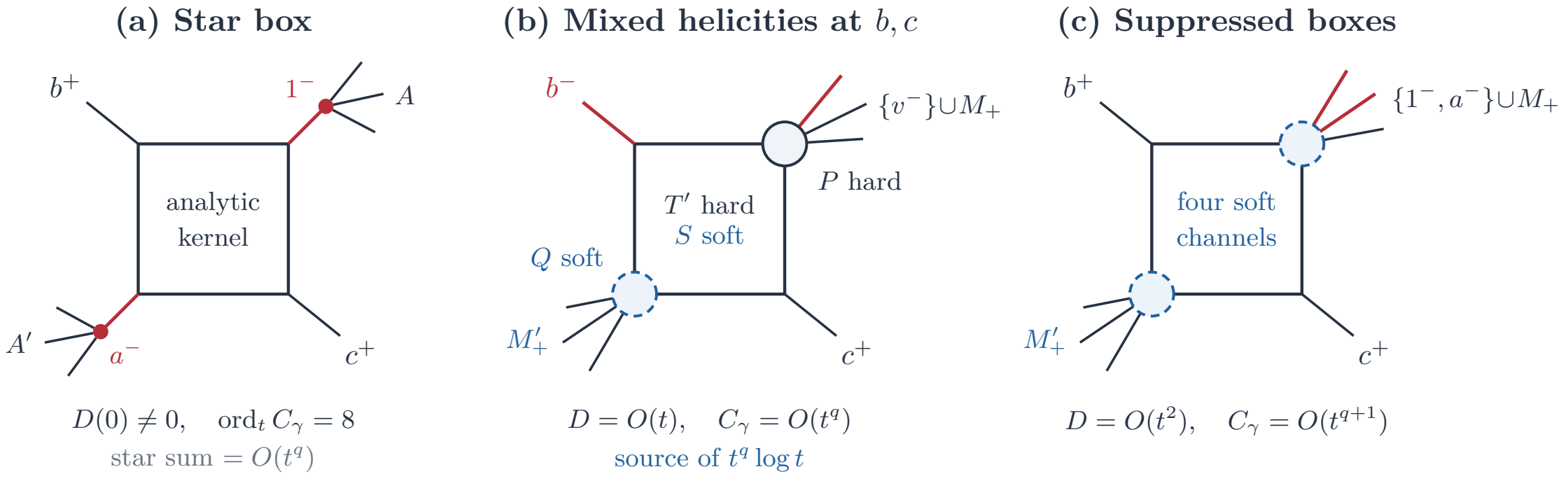

\begin{theorem}[Cancellation below the tree power]
\label{one:low}
For $n\ge6$ on the regular paths above, all coefficients at powers
$t^8,\ldots,t^{n+2}$ of the one-loop finite part $F_1$ vanish.
\end{theorem}
Only the stars in figure~\ref{fig:one-loop-boxes}(a) can contribute
below $t^q$, because their endpoint trees use exclusively mixed-helicity
edges. Their scalar kernels are analytic, so one can sum the low powers
coefficient by coefficient. The low-order logarithmic terms cancel by
Lagrange interpolation in the nodes of $b,c$ at fixed channel, and the
corresponding dilogarithm coefficients cancel by an alternating sum
over corner assignments at fixed $b,c$.
\begin{proof}
The non-star coefficients obey the bounds just derived, and a scalar
kernel grows at most quadratically in $\log t$. In a star box,
$u=O(t)$. Euler reflection and
$(1-uS)(1-uT')=(1-uP)(1-uQ)$ express its kernel as the signed sum of
\begin{equation}
 -\tfrac12 L_X^2+L_X\log(1-uX)+\operatorname{Li}_2(uX).
 \label{one:analytic}
\end{equation}
The four dilogarithm constants cancel. A singleton entry in this
rewritten sum gives zero. The nonzero entries are hard, and the
boundary values in \eqref{one:analytic} are analytic along the path.
We remove $\langle1a\rangle^8$ for the next two sums.

\paragraph{Interpolation at a fixed channel.}
Pass to an analytic frame $\lambda_1=(1,0)$, $\lambda_a=(r,\delta)$,
$\lambda_i=(x_i,1)$, $\widetilde\lambda_i=s_i(1,y_i)$ for $i\in P_+$.
Here $r(0)\ne0$, $\delta,y_i=O(t)$ and $s_i(0)\ne0$. For $v=1,a$ and
a positive set $A$, root the tree at $v$ and divide each undeleted row
of its Laplacian by the weight of the edge to $v$. This gives
\begin{align}
 h(b,\{v\}\cup A,c)&=w_v^{|A|-2}\prod_{i\in A}\frac{g_{vi}}{w_i}
     \mathcal F_v(A),&\quad \mathcal F_v(A)&=\det(1+K_v),\nonumber\\
 (K_v)_{ij}&=-\frac{w_jg_{ij}}{w_vg_{vi}}\quad(i\ne j),&
 (K_v)_{ii}&=-\sum_{j\ne i}(K_v)_{ij}.\label{one:forest}
\end{align}
Fix a partition $I=\{1\}\cup A$, $I^c=\{a\}\cup A'$ and the entry
$X=s_I$. Use $H_{ij}=\sum_{k\in I}\langle ik\rangle[kj]$ and
$\beta_i=\prod_{j\in P_+\setminus\{i\}}(x_i-x_j)^{-1}$. The
two-dimensional momentum determinant gives
$D=H_{bc}H_{cb}$ and $1-uX=H_{bb}H_{cc}/D$. For each $b,c$ delete
them from $I,I^c$ to form $M,M'$. Extending the forests back to the
full positive sets $A,A'$ gives the signed weight
\begin{multline}
 \sigma_XDhh=(-1)^N\Big(\prod_{i\in A}g_{1i}\prod_{i\in A'}g_{ai}\Big)
     \beta_b\beta_c(x_b-x_c)^2\\ {}
     \times w_1^{|A|-2}w_a^{|A'|-2}\mathcal F_1(A)\mathcal F_a(A').
 \label{one:fixed}
\end{multline}
Here $hh$ denotes the two half-soft factors. For instance, $w_b=0$
makes column $b$ diagonal in the extended determinant. On side $I$
its diagonal times $w_1g_{1b}$ equals $H_{cb}$. On side $I^c$ the
corresponding product equals $-H_{cb}$. The second deleted column
gives the factor involving $H_{bc}$. The product over all remaining
$1/w_i$ is $(-1)^{N-1}\beta_b\beta_c(x_b-x_c)^2$.

Replace $(x_b,x_c)$ by independent $(x,y)$ in the second line of
\eqref{one:fixed}. Then $w_i=(x-x_i)(x_i-y)$, $w_1=-1$ and
$w_a=-(r-\delta x)(r-\delta y)$. Each insertion of $K_v$ costs a
power of $t$ through $g_{ij}$ and adds at most one degree in each
variable. Integer powers of $w_a$ have the same degree bound order by
order. Thus the relative $t^d$ coefficient has bidegree at most $(d,d)$.
Taking the highest coefficient of the Lagrange interpolant gives
$\sum_i\beta_i x_i^h=0$ for $0\le h\le N-2$, also for the moving
nodes at each $t$. The symmetric pair sum is half the ordered double
sum. Through $d=N-3$, each monomial after multiplication by
$(x-y)^2$ has degree at most $N-2$ in at least one variable. Thus the
signed star weight of \emph{each} channel is $O(t^{q+1})$ with the
helicity factor restored. This treats the $L_X^2$ terms.

For $\log(1-uX)$ use the sum of $\log|H_{bb}|,\log|H_{cc}|$ minus
$\log|H_{bc}|,\log|H_{cb}|$. The factors $s_b,s_c$ cancel in this
combination. Each diagonal logarithm leaves one interpolation variable
free. At fixed $c$, $H_{bc}/s_c$ is affine in $x$, with a nonzero
constant at $t=0$ and slope $O(t)$ owing to positive square collinearity.
The logarithm has degree at most $d$ in $x$ at relative order $d$.
The same argument with $b,c$ interchanged treats $H_{cb}$. Each
summand, including $(x-y)^2$, leaves one interpolated variable of
degree at most $d+2$. It vanishes after summation for $d\le N-4$.

\paragraph{Alternation at fixed massless corners.}
Now fix $b,c$ and assign the other $N-2$ positive labels to the two
roots with indicators $e_i=0,1$, taking $e_i=1$ on the $1$ side.
Extract $(-1)^{\sum e_i}$ from the half-soft product. We show that its
relative $t^d$ coefficient is a multilinear polynomial of degree at
most $d$ in the indicators. Each directed edge from $i$ to $j$ carries
$w_jg_{ij}/(w_vg_{vi})$ for colour $v$. In \eqref{one:forest} the individual left
to right ratios satisfy $w_1g_{1i}/(w_ag_{ai})=-1+O(t)$ by momentum
conservation. Their leading sign has just been removed. An uncoloured
forest component on $k$ vertices, oriented towards its attaching
vertex, contains $k-1$ internal edges. Its first possible order is
$k-1$, and the two colour assignments at this order give a common
factor times
\[
       \prod_{i\in G}e_i+(-1)^{k-1}\prod_{i\in G}(1-e_i).
\]
The degree-$k$ term cancels. Higher orders have degree at most $k$,
and multiplying components preserves the order bound.

Apply this count to the series for $D\operatorname{Li}_2(uX)$. Each
$H_{bc}$ or $H_{cb}$ has a nonzero indicator-independent constant, and
its indicator-dependent summands are linear and $O(t)$. Consequently
$D^{1-j}$ satisfies the same degree bound for $j\ge1$. The order-$d$
part of $X$ has degree at most $d+1$, since mixed pairs contribute
linearly and positive pairs are quadratic and cost one power of $t$.
In $D(uX)^j=(-s_{bc})^jD^{1-j}X^j$ the factor $s_{bc}^j=O(t^j)$
pays for the $j$ extra degrees. The resulting order-$d$ polynomial
has degree at most $d$. Its sum against $(-1)^{\sum e_i}$ vanishes
for $d<N-2$. Equations~\eqref{one:fixed} and \eqref{one:analytic}
therefore cancel every star term through $t^{8+N-4}=t^{n+2}$.
\end{proof}

\subsection{A channel identity at fixed kinematics}

The gravitational one-loop infrared pole is proportional to
$\mathcal T\sum_{i<j}s_{ij}L_{s_{ij}}/\epsilon$~\cite{Weinberg,NS}.
In the scalar-box basis each nonzero entry $X$ contributes
$-\sigma_X C_\gamma(-X-i0)^{-\epsilon}/\epsilon^2$, whose
single pole is $\sigma_X C_\gamma L_X/\epsilon$. The box weights
must therefore cancel logarithms of multiparticle invariants and
reproduce the tree coefficient of each two-particle logarithm.
Six-point examples of these infrared relations appear in
ref.~\cite{IRAmplitude}. Here the half-soft representation gives a
rational identity for each partition. Its proof also fixes the weights
of the squared logarithms in $F_1$.

For a physical partition $I|I^c$ with at least two labels on each
side, let $W_{I|I^c}$ sum $\sigma_X C_\gamma$ over all its occurrences
in \eqref{one:kernel}. At fixed generic kinematics,
\begin{equation}
 W_{I|I^c}=\mathcal T\sum_{\substack{K\in\{I,I^c\}\\|K|=2}}s_K .
 \label{one:residues}
\end{equation}
The identity is rational and determines the channel weights before the
helicity-zero limit.

\begin{proof}
Write $\lambda_i=(z_i,1)$ at fixed $t$ and replace the two reference
spinors in \eqref{one:halfsoft} by $(x,1),(y,1)$. In the rational
differential
\[
 (x-y)^2 h(x,I,y)h(x,I^c,y)\,dx\,dy,
\]
a vertex of degree $d$ contributes $[(x-z_i)(z_i-y)]^{d-2}$ to its
tree. Only leaves have poles. Each half-soft function is $O(x^{-2})$
as $x\to\infty$ at fixed $y$ by the sum of tree degrees, and the
same holds in $y$. The ordered double residues thus sum to zero.
Diagonal double residues vanish. At distinct nodes $b,c$ the two leaf
residues times $(z_b-z_c)^2$ give $-1$. Unless the leaves constitute
a whole side, delete them and sum independently over their possible
neighbours. On side $I$ their attachment sums are
\[
 \sum_{j\in I\setminus\{b,c\}}w_jg_{bj}=H_{cb},\qquad
 \sum_{j\in I\setminus\{b,c\}}w_jg_{cj}=H_{bc},
\]
where the identity for a deleted leaf is used on the side to which it
belongs. The complementary side reverses each sign by conservation.
The resulting residue is $\sigma_XDhh$, with positive sign for
leaves on opposite sides and negative sign for two leaves on one side.

For $I=\{b,c\}$ the only tree has weight $g_{bc}/(w_bw_c)$. Its residue is $-s_{bc}h(b,I^c,c)/\langle bc\rangle^2$.
The quotient here equals $\mathcal T/\langle1a\rangle^8$. To see
reference independence directly, use the symmetric Laplacian $L$ with
off-diagonals $-[ij]/(z_i-z_j)$. Conservation annihilates the vectors
$1,z,z^2$. In particular $Lw=0$ for $w_j=(z_b-z_j)(z_j-z_c)$.
The weighted Laplacian on $I^c$ is the restriction of
$\operatorname{diag}(w)L\operatorname{diag}(w)$. For any $r\ne b,c$,
\[
 \frac{h(b,I^c,c)}{\langle bc\rangle^2}=
 \frac{\det L_{\widehat{bcr},\widehat{bcr}}}
 {[(z_b-z_c)(z_c-z_r)(z_r-z_b)]^2}.
\]
The minors of a symmetric corank-three matrix are proportional, at this
size, to the squares of its complementary kernel minors. Here the
kernel minors are the displayed Vandermonde determinants. At larger
corank all these minors vanish. Choosing $b=1,c=a$ in the quotient
therefore identifies it with the stripped tree. The same extra term
occurs if $I^c=\{b,c\}$. Summing residues, dividing by two for the
ordered pairs, and restoring the helicity factor proves
\eqref{one:residues}.
\end{proof}

Equation~\eqref{one:residues} supplies a continuous prescription on
paths with extra degeneracies. First group squared logarithms by
partition, leaving $-\tfrac12\mathcal T\sum_{i<j}s_{ij}L_{s_{ij}}^2$.
Each $sL_s^2$ extends continuously to zero. A massless dilogarithm
entry in \eqref{one:kernel} retains its value $\operatorname{Li}_2(1)$.
At fixed $t>0$ with regular angle denominators,
$D\operatorname{Li}_2(1-u(X+i0))$ tends to zero for $D\to0$, since
the boundary values grow at most as $1+|\log|D||^2$.

\subsection{Channel selection on the Feynman sheet}

\begin{theorem}[First possible power and its logarithm]
\label{one:branch}
On a regular real transverse path,
\begin{equation}
 F_1(t)=t^q\left[i\pi T_0\sum_{j\in P_+}|s_{1j}(0)|\log t+B
       +O\!\left(t(1+|\log t|^2)\right)\right].
 \label{one:expansion}
\end{equation}
The constant $B$ is finite and path dependent. If $T_0\ne0$,
$F_1/\mathcal T$ grows logarithmically in magnitude.
\end{theorem}
\begin{proof}
The star kernels are analytic, and their signed weights in each
physical channel vanish through $t^q$ by \eqref{one:fixed}. Only boxes with one negative massless corner $b$ can contribute a
logarithm at $t^q$. Let $c$ be their positive massless corner, $v$
the other negative leg in $M$, and $M'$ the all-positive corner.
Here $S=s_{M'\cup\{c\}}$ and $Q=s_{M'}$ are soft, whereas
$T'=s_{M'\cup\{b\}}$ is hard, and $P=s_M$ is hard except when
$M=\{v\}$, in which case it vanishes identically. For $c_\gamma=[t^q]C_\gamma\ne0$, the half-soft
bounds force $D$ to have order one, and $u=u_0/t+O(1)$.

Write $X=x_0t^d(1+O(t))$ for a nonzero channel, with
$H(x)=\log|x|-i\pi\mathbf1_{x>0}$ and
$U(u)=\log|u|+i\pi\mathbf1_{u>0}$. Before multiplication by
$\sigma_X$, the logarithmic coefficients from this position in
\eqref{one:kernel} are
\[
\renewcommand{\arraystretch}{1.2}
\begin{array}{c@{\quad}|@{\quad}c@{\qquad}c}
 & [\log^2 t] & [\log t] \\ \hline
 d=0 & \tfrac12 & -H(x_0)-U(u_0)\\
 d\ge1 & -d^2/2 & -d\,H(x_0)
\end{array}
\]
For $d=0$ the dilogarithm inversion at infinity gives the extra
logarithms with the indicated $i0$. For $d\ge1$, $uX$ is finite and
the dilogarithm contributes a constant at this order. An identically
massless entry also contributes only a constant under the continuous
prescription.

In each soft channel \eqref{one:residues} has zero coefficient at
$t^q$. Since $d,x_0$ depend only on the channel, both soft logarithms
cancel. The hard double logarithms sum to zero by
$s_{aj}(0)=-s_{1j}(0)$. The two hard entries in a box have opposite
signs, so their $U$ terms cancel except for $M=\{v\}$. There the sole
hard sign is positive, $S=s_{1a}$, $D=ST'$, and
$T'(0)=-s_{bc}(0)$. Thus all such boxes have the same
$u_0=1/\partial_ts_{1a}|_0$. They are exactly the occurrences of the
soft partition $\{1,a\}|P_+$ in the class contributing at order $q$,
all with positive sign. Equation~\eqref{one:residues} makes their
combined $U$ term zero. With $b_j=s_{1j}(0)$, the surviving hard
pair terms are $T_0\sum_j[-b_jH(b_j)+b_jH(-b_j)]$. Each pair gives
$i\pi T_0|b_j|$. Expanding the analytic spinors and the weight-two
kernel one order further gives the remainder in \eqref{one:expansion}.
\end{proof}

The two-particle hard logarithms combine into an extremal invariant.
Indeed, $s_{\{1\}\cup A}(0)=\sum_{j\in A}s_{1j}(0)$ for every
$A\subseteq P_+$ because all positive-pair invariants vanish.
Since the mixed invariants are nonzero and sum to zero, the unique
maximizer is $J=\{j\in P_+:s_{1j}(0)>0\}$, and
\begin{equation}
          \sum_{j\in P_+}|s_{1j}(0)|=2s_{\{1\}\cup J}(0).
 \label{one:selected}
\end{equation}
Both sides of this partition contain at least two legs. The
imaginary logarithm in \eqref{one:expansion} comes from the Feynman
phases of the hard pairs $(1,j)$ and $(a,j)$. The partition
$\{1\}\cup J$ records their signed sum and can be multiparticle even
when its own weight in \eqref{one:residues} vanishes. Within a real
sign region $J$ is fixed, so the expression on the right continues
analytically from that region.

\subsection{A positive family and direct box sums}
\label{one:examples}

The coefficient in \eqref{one:expansion} is nonzero for every member
of an explicit family. Choose $m,k\ge2$, $n=m+k+2$, $a=m+2$. In an
oriented basis $(A,U,B_0,V)$ put
\begin{align}
 X(r)&=A+rU+r^2B_0,& Y(s)&=B_0+sV-s^2A,\nonumber\\
 Z_0&=(A,X_1,\ldots,X_m,B_0,Y_1,\ldots,Y_k),&Z(t)&=Z_0\exp(tE),
 \label{one:uniform}
\end{align}
where $X_i=X(r_i)$ and $Y_j=Y(s_j)$ are uniformly spaced in their
\emph{parameters} on $[1/4,1/2]$ and $[1/3,2/3]$ respectively. The
matrix $E_{ij}=\mathbf1_{|i-j|=1}$ is the adjacency of the open chain.
Read $u=(u_1,u_2,u_3,u_4)$ as
$(\lambda^1,\lambda^2,\mu^{\dot1},\mu^{\dot2})=(u_1+u_3,u_2+u_4,u_1,u_2)$.
In cyclic order the momentum-twistor incidence relation~\cite{HodgesTwistors} gives
\begin{equation}
 \widetilde\lambda_i=
 \frac{\mu_{i-1}\langle i,i+1\rangle+\mu_i\langle i+1,i-1\rangle
                +\mu_{i+1}\langle i-1,i\rangle}
 {\langle i-1,i\rangle\langle i,i+1\rangle}.
 \label{one:incidence}
\end{equation}
All ordered four-minors of $Z_0$ are nonnegative. A three-to-one split
between the arcs gives a positive Vandermonde times the opposite
parameter. A two-to-two split gives $(r'-r)(s'-s)(1+rr'ss')$.
The end columns are obtained by limits, and the four-column
determinant satisfies $[A,X(r),B_0,Y(s)]=rs>0$. The generator induced by $E$
on ordered exterior powers moves one index to a free neighbour with
coefficient $+1$. These moves connect all ordered index sets of fixed
size, so every ordered minor of $\exp(tE)$ is positive for $t>0$.
Cauchy--Binet then gives strict positivity of the path.

To evaluate the tree cofactor, rescale each positive angle to $(x_i,1)$
and compensate on its square spinor. Denote the first dotted component
at zero by $q_i$ and the derivative of the second by $p_i$. The nodes
are $x_X=r+1/r$, $x_Y=1/s-s$. Each list is strictly decreasing. A
cross-list equality would give $rs(r+s)=r-s$. With $v=r-s$ and
$w=r+s$ this implies $(wv+2)^2=w^4+4$, a rational right triangle
with legs $w^2,2$ and square area, excluded by Fermat's right-triangle
theorem (no rational right triangle has square area). Thus all nodes are distinct.

Equation~\eqref{one:incidence} makes $q_i$ the previous minus the
next edge slope of $\mu^{\dot1}/\lambda^2$ versus $x$. The interior
slopes are $1/(1-rr')$ and $ss'/(1+ss')$ on the two arcs. At $A,B_0$
the edge slopes are $1,0$. Thus $q_h>0$ for the sole node $h=X_m$,
with $q_h=1/(1-r_{m-1}r_m)$, and $q_i<0$ otherwise. At the endpoint
the negative dotted spinors are $(5/4,1)$ and $(-7/3,-1)$. They give
$[1a](0)=13/12$, $s_{1i}(0)=-q_i$, $\sum q_i=0$ and
$\sum x_iq_i=13/12$. Every proper nonempty subset sum of the $q_i$
is nonzero. Differentiating \eqref{one:incidence} using $Z'_0=Z_0E$
gives $\delta_1=7/12$ and three nonzero $p_i$,
\begin{equation}
  p_{X_m}=1/3,\qquad p_{Y_1}=1/2,\qquad p_{Y_k}=-1/4.
 \label{one:three}
\end{equation}
Indeed $\mu^{\dot2}-\lambda^2=0$ along the first arc and
$\mu^{\dot2}=0$ along the second. Their first-order mismatches on the
shared endpoints are the only ones entering these three reconstructions.

In the affine frame $L_*$ has edges $(q_ip_j-p_iq_j)/(x_i-x_j)$.
Put $P_x(z)=\prod_i(z-x_i)$,
$A_q=P_x\sum_iq_i/(z-x_i)$ and $A_p=P_x\sum_ip_i/(z-x_i)$. By
\eqref{one:three},
\[
 \frac{A_p}{P_x}=\frac{1/3}{z-5/2}+\frac{1/2}{z-8/3}-\frac{1/4}{z-5/6}.
\]
The zeros of $A_p$ are the inactive nodes and the two irrational
conjugate roots of $126z^2-333z+25$. At any external node
$A_q(x_i)=q_iP_x'(x_i)\ne0$. For the smaller quadratic root
$\alpha<5/6\le x_i$, the positive weights $-q_i/q_h$ on $i\ne h$
sum to one and have mean $x_h-(13/12)/q_h<x_h$. Strict decrease and
convexity of $x\mapsto1/(x-\alpha)$ imply
$\sum_{i\ne h}(-q_i/q_h)/(x_i-\alpha)>1/(x_h-\alpha)$, hence
$A_q(\alpha)\ne0$. Rationality excludes the conjugate root as well.

For any $L_*f=0$, let $A_{qf},A_{pf}$ denote the same polynomials
with $q_i,p_i$ multiplied by $f_i$. The rational function
$(A_qA_{pf}-A_pA_{qf})/P_x^2$ vanishes at infinity and has only
simple poles with residues $-(L_*f)_i=0$. Coprimality therefore
forces $A_p\mid A_{pf}$. Since $\deg A_p=N-1$ and
$\deg A_{pf}\le N-1$, both weighted polynomials are common constant
multiples of their unweighted counterparts. Evaluating $A_{qf}$ at
each $x_i$ makes $f_i$ constant. Thus $L_*$ has corank one and
\begin{equation}
  T_0\ne0,\qquad 2s_{\{1\}\cup J}(0)=2q_h=\frac{16(m-1)}{6m-5}
 \label{one:uniformresult}
\end{equation}
for every $m,k\ge2$.

For the second six-point test, use the standard basis of $\mathbb R^4$
as $z_2,z_3,z_5,z_6$ in that order. Set
\begin{align*}
 z_4&=(-z_2+(1+t)z_3+4z_5-(4+2t)z_6)/3,\\
 z_1&=z_2-z_3+z_4-z_5+z_6,\qquad
 Z_i=\begin{pmatrix}2&3&5&8\\1&1&1&1\\-1&-4&-2&-7\\0&0&1&1\end{pmatrix}z_i .
\end{align*}
Read $Z_i=(\lambda_i,\mu_i)$ in \eqref{one:incidence}. In the $z$
frame an ordered maximal minor complementary to $i<j$ equals
$(a_j-a_i)/3$ with $a=(0,1,1+t,3,4,4+2t)$, proving strict positivity
for small $t>0$ after the positive-determinant change of frame.
In both six-point examples the negative pair is $1,4$, and the set
$J=\{2,5,6\}$ selects $s_{34}$ by complementation. Summing all 105
boxes in the original twistor spin frames gives the coefficients below.
The affine rescalings used to prove nonvanishing are undone here.
\begin{center}
\setlength{\tabcolsep}{10pt}
\begin{tabular}{lrrr}
\toprule
Path & $T_0$ & $2s_{34}(0)$ & $[t^9\log t]F_1/(i\pi)$ \\
\midrule
Polynomial & $-176128/277950234375$ & $32/11$ & $-5636096/3057452578125$ \\
Exponential & $1857963429/471295$ & $16/7$ & $4246773552/471295$ \\
\bottomrule
\end{tabular}
\end{center}
At $n=10$, $(m,k)=(4,4)$, the 5715-box sum gives
\[
 \frac{[t^{13}\log t]F_1}{i\pi}=
 \frac{83850614486052722739896746579680892734}{68826658707171087771872866796875}.
\]
The exact box expansion checks the vanishing of each lower power,
the double logarithm and the real single logarithm independently
of \eqref{one:cofactor}. Direct complex ball evaluations of the ten-point
sum at $t=10^{-16},10^{-24}$, using 400 decimal digits, give a slope
of $F_1/t^{13}$ against $\log t$ within $1.7\times10^{-14}$ in relative
error of the predicted coefficient.

\section{Lorentzian tree zeros and the one-loop absorptive part}
\label{sec:physical}

\subsection{Five gravitons at a regular tree zero}

Consider a planar \(2\to3\) process with
negative-helicity gravitons on the two incoming beams, denoted by \(A,a\).
Write each momentum in the all-outgoing convention as
\(p_i=q_i v_i v_i^\dagger\), with \(v(z)=(1,z)^{\mathsf T}\) and
\(v(\infty)=(0,1)^{\mathsf T}\).
The two incoming beams have negative energy in this convention.
Choose the real nodes and signed weights
\begin{equation}
 \begin{array}{c|cc|rrr}
 i&A&a&3&4&5\\ \hline
 z_i&0&\infty&-1&1/2&2\\
 q_i&-1&-1&4/9&4/9&1/9
 \end{array}.
 \label{phys:point}
\end{equation}
The outgoing weights obey \(\sum q_i=\sum q_i z_i^2=1\) and
\(\sum q_i z_i=0\), so their momentum sum is the identity matrix,
cancelled by the two beams. The incoming invariant is \(s_{Aa}=1\).
The ten two-particle invariants are \(1,4/9,1/9,-4/9,-1/9\) with
multiplicities \(1,2,1,4,2\). At five points they exhaust the physical
channels up to complementation.

Use physical spinors
\(\lambda_i=\sqrt{|q_i|}\,v_i\) and
\(\widetilde\lambda_i=\operatorname{sgn}(q_i)\bar\lambda_i\).
Then \(\langle Aa\rangle=1\) and
\([ij]/\langle ij\rangle=1\) for every outgoing pair.
The three positive vertices in the MHV tree have weights
\(w_i=\langle Ai\rangle\langle ia\rangle=q_i z_i\), namely
\((-4/9,2/9,2/9)\). Adding the three spanning trees over a common
denominator leaves \(\sum_i w_i=\sum_i q_i z_i\) in the numerator.
This is the off-diagonal component that vanishes by momentum conservation.
For the same momenta the tree and the one-loop absorptive part are
\begin{equation}
 \mathcal T=\frac{w_3+w_4+w_5}{w_3 w_4 w_5}=0,\qquad
 \operatorname{Im}F_1^{\rm phys}=-81\pi\log 2 .
 \label{phys:five}
\end{equation}
The logarithm comes from the unequal magnitudes \(4/9\) and \(2/9\)
of the weights. In the subset formula~\eqref{ir:forward} of
appendix~\ref{app:lorentz}, set \(f(y)=y\log|y|\), \(f(0)=0\).
The surviving terms in \(\operatorname{Im}F_1^{\rm phys}/\pi\) are
\[
 \frac{729}{8}\left[2f(-4/9)+4f(2/9)\right]=-81\log 2 .
\]

Summing the discontinuities of the 30 one-mass boxes on the \(-i0\)
sheet gives an independent check.
In the notation of~\eqref{one:boxkin}, all
30 Gram determinants are nonzero and \(u<0\). A channel position
\(X>0\) contributes \(C_\gamma\sigma_X[\log X-\log(1-uX)]\) to
\(\operatorname{Im}F_1/\pi\) at \(\mu^2=1\). The first log comes from
the squared logarithm and the second from the dilogarithm.
Using the rational spinors above in~\eqref{one:boxes}, the first terms
sum to zero. Factoring the arguments of the second terms into primes
leaves coefficient \(-81\) for \(\log 2\) and zero for every other prime
logarithm. The supplementary check \texttt{physical\_five\_point.py}
performs this sum with rational arithmetic.

At this point the tree zero removes the one-loop infrared pole,
so the one-loop amplitude itself is finite. Since
\(\operatorname{Im}\sigma=-2\pi\), the two-loop simple pole
\(\sigma F_1\) survives by~\eqref{ir:poles}.

\subsection{The two-incoming sign theorem}

At a regular point of Lorentzian kinematics the null directions are
distinct and every partition with at least two legs on each side has a
nonzero invariant. Planarity means that in the centre-of-mass frame
all spatial momenta lie in a plane containing the beams. The
MHV gravity divisor is the vanishing locus of the tree
amplitude~\cite{Hodges,KOPTZ,MZ,KVY2}.

\begin{theorem}[Two incoming gravitons]
\label{phys:two-incoming}
Let \(n\ge5\) and take regular Lorentzian kinematics with exactly two
incoming momenta and positive outgoing energies. The MHV graviton tree
vanishes if and only if the scattering is planar. At every such tree
zero the one-loop amplitude is infrared finite and nonzero for every
choice of the two negative-helicity gravitons. In the real planar spin
frame \(\lambda_i=\sqrt{|q_i|}\,v_i\), its absorptive part is strictly
negative for every such choice.
\end{theorem}

\begin{proof}
Planar gravity zeros also appear in tree-level analyses~\cite{Planar}.
For the converse, appendix~\ref{app:lorentz} uses positivity of the
outgoing energies to show that a common root of the two momentum
interpolation polynomials forces all outgoing nodes onto a straight
line through the origin of the celestial chart. For planar scattering,
rotate the finite nodes to real \(x_i\). The two interpolants then
become \(U(w),wU(w)\) with a common factor of degree \(n-4\), so their
resultant vanishes.

Assign the negative helicities to the beams \(A,a\). Write \(N=n-2\),
let \(s_{\rm in}>0\) be the incoming-pair invariant, and set
\(y_i=q_ix_i\) for each outgoing particle. If \(\rho\) is the density of
the sum of independent centred uniform variables of widths \(|y_i|\),
the complete box sum in appendix~\ref{app:lorentz} gives
\begin{equation}
 -\frac{\operatorname{Im}F_1^{\rm phys}}{\pi}
 =4(N-2)!\,s_{\rm in}^3
       \int_0^\infty\frac{-\rho'(u)}{u}\,du>0 .
 \label{ir:sign}
\end{equation}
The density is even, is nonincreasing for \(u>0\) and decreases somewhere
on its compact support, which gives the strict inequality.
For either the tree or the one-loop finite part \(M\),
the supersymmetry Ward identity~\cite{SWI} gives
\begin{equation}
 M_{r^-s^-}=\frac{\langle rs\rangle^8}{\langle Aa\rangle^8}
          M_{A^-a^-} .
 \label{phys:ward}
\end{equation}
In this frame the angle brackets of distinct legs are real and nonzero.
Their ratio to the eighth power is strictly positive, so every MHV
assignment has the sign in~\eqref{ir:sign}. Nonvanishing is independent
of the external spinor phases.
The only one-loop infrared pole has residue \(\sigma\mathcal T\), which
vanishes at the tree zero.
\end{proof}

Since \(\operatorname{Im}\sigma=-2\pi s_{\rm in}\), the next-to-highest
infrared pole survives at every loop order \(L\ge2\) on this locus,
while the highest pole vanishes (section~\ref{sec:ir}).

\subsection{Common zeros at other energy signatures}

With four incoming gravitons and three outgoing, a discrete symmetry
can force the tree and one-loop amplitudes to vanish together even at a
nonplanar point. To see this, put
\(\omega=e^{2\pi i/3}\), \(\lambda_i=(1,z_i)^{\mathsf T}\),
\(\widetilde\lambda_i=q_i(1,\bar z_i)^{\mathsf T}\), and take
\[
 \begin{array}{c|ccc}
      z_i & 0 & \omega^j & 2\omega^j\\
      q_i & -9 & 4 & -1
 \end{array}
 \qquad(j=0,1,2\ \text{on each triangle}).
\]
The two triangles have vanishing first moment, and their weights cancel
both diagonal entries of the momentum sum after inclusion of the central
leg. None of the 56 physical partitions has zero invariant. The distinct
two-particle values are \(-36,-28,-4,12,36,48\), and the three-particle
values are \(-44,-28,-24,-20,-8,-4,16,36,84,144\). A simultaneous rotation
of the triangles is a permutation preserving every transported scalar
kernel and its boundary prescription. Under this permutation angle and
square brackets acquire factors \(\omega\) and \(\omega^{-1}\),
respectively. A half-soft tree on \(k\) legs has weight
\(\omega^{-2k-2}\) by~\eqref{one:halfsoft}. After stripping
\(\langle1a\rangle^8\), the one-loop box coefficient therefore has
weight \(\omega^{-2n}=\omega^{-2}\ne1\). The stripped tree,
\(h(1,P_+,a)/\langle1a\rangle^2\), has the same weight and is
permutation symmetric by the Hodges formula. Both amplitudes must
vanish. The exact calculation at this point groups 315 boxes into 63
distinct kernels with zero total coefficient in \(\mathbb Q(\omega)\).
All box Gram determinants are nonzero.

One can also vary \(F_1\) at fixed tree corank and energy signature.
Here excess corank denotes the nullity remaining after removal of
the three universal Hodges null directions.
Consider a second seven-point configuration with nodes
\(0,-3\omega^j,8\omega^j\) and respective weights \(-165,64,-9\).
Here the interpolation map is \(f(z)=49z^2/(z^3-120)\).
The seven directions are simple solutions of \(\bar z=f(z)\).
Their two degree-four binary interpolants are proportional to
\((z^3-120,49z^2)\) in the affine chart and share precisely the root at
infinity. The excess Hodges corank is one. The same rotation enforces
\(F_1=0\).

Translate \(f\) to \(f+t\eta\), keeping the central
weight fixed and solving
\(\sum_iq_i z_i^k/(z_i^3-120)=0\) for \(k=0,\ldots,4\).
The seven simple directions persist. These conditions preserve
conservation and the cubic reduced interpolation map. Six independent
real rows fix the other six weights. The rank remains six nearby since
the complex row space and its conjugate share the four functions
\(1,z,\bar z,|z|^2\). Along these deformations the tree stays at corank
one. Let \(D(\eta)\) be the derivative of the stripped amplitude
\(\widehat F_1=F_1/\langle1a\rangle^8\) along \(f+t\eta\) at \(t=0\).
Since \(f(\omega z)=\omega^2 f(z)\), rotating the nodes sends the
translation direction \(1\) to \(\omega^2\). The stripped amplitude
acquires the phase \(\omega^{-14}=\omega\), hence
\[
 D(\omega^2)=\omega D(1).
\]
Since \(\omega\) is nonreal, the two images are linearly independent
over \(\mathbb R\) as soon as \(D(1)\ne0\). Differentiating all 315 boxes
certifies \(D(1)\ne0\). In the ordered directions \(1,\omega^2\), the
real Jacobian determinant \(\operatorname{Im}[\overline{D(1)}D(\omega^2)]\) lies in
\((1.35964318,1.35964320)\times10^{26}\).
The interval calculation, including the rank test, is supplied in
\texttt{hodges\_lorentzian\_seven\_lens.py}. At distinct Lorentzian
directions, each pair of energies of the same sign gives a positive
\(s_{ij}\). Hence \(\operatorname{Im}\sigma=-\pi\sum_{s_{ij}>0}s_{ij}<0\)
on this seven-point family. The nonzero Jacobian makes \(F_1\ne0\)
on a punctured neighbourhood within the real two-parameter deformation.
Thus the next-to-highest infrared pole takes both zero and nonzero
values at kinematically regular points of the same excess-corank-one
stratum.

\section{Infrared poles and two-loop scale dependence}
\label{sec:ir}

\subsection{The two highest poles}

At a regular tree zero in real two-incoming scattering, the one-loop
amplitude is finite and nonzero by section~\ref{sec:physical}. Its value enters the infrared poles
at higher loops through the gravitational soft exponent, which is
one-loop exact~\cite{Weinberg,NS}. In the normalization of
section~\ref{sec:one}, the amplitude factorizes as
\begin{equation}
 M(g,\epsilon)=\exp(g\,\sigma/\epsilon)\,\mathcal H(g,\epsilon),
 \qquad
 \sigma=\sum_{i<j}s_{ij}\log\frac{-s_{ij}-i0}{\mu^2},
 \label{ir:exp}
\end{equation}
where $g$ counts loops, $\mathcal H^{(0)}=\mathcal T$, and the hard
function $\mathcal H$ is finite at $\epsilon=0$ once any independent
ultraviolet poles are accounted for. Because $\sum_{i<j}s_{ij}=0$ by
momentum conservation, $\sigma$ is independent of the scale $\mu$. With a
four-dimensional external tree there is no evanescent tree piece
multiplying the poles, so the one-loop pole is $\sigma\mathcal T/\epsilon$
and $F_1=[\epsilon^0]\mathcal H^{(1)}$. Expanding the exponential gives,
for every $L\ge2$,
\begin{equation}
 [\epsilon^{-L}]M^{(L)}=\frac{\sigma^L}{L!}\,\mathcal T,\qquad
 [\epsilon^{1-L}]M^{(L)}
 =\frac{\sigma^{L-1}}{(L-1)!}\,F_1 .
 \label{ir:poles}
\end{equation}
At a regular planar $2\to n-2$ Lorentzian zero,
$\mathcal T=0$ removes the highest pole
$[\epsilon^{-L}]M^{(L)}$ at every loop order, whereas
$\operatorname{Im}\sigma=-2\pi s_{\rm in}\ne0$ and $F_1\ne0$ ensure that
the next-to-highest pole $[\epsilon^{1-L}]M^{(L)}$ survives for all
$L\ge2$.

On the split-signature transverse paths of section~\ref{sec:one}, every
soft pair contributes $s_{ij}\log(-s_{ij}-i0)=O(t(1+|\log t|))$, while
each hard pair has opposite endpoint invariants
$b_j=s_{1j}(0)=-s_{aj}(0)$. The real logarithms cancel pairwise, giving
\[
 \sigma=\sigma_0+O(t(1+|\log t|)),\qquad
 \sigma_0=-i\pi\sum_{j\in P_+}|b_j|\ne0 .
\]
Inserting the one-loop expansion~\eqref{one:expansion} into~\eqref{ir:poles}
gives the two-loop poles at the reference scale,
\begin{align}
 [\epsilon^{-2}]M^{(2)}&=t^q\big(\tfrac12\sigma_0^2T_0+o(1)\big),\\
 [\epsilon^{-1}]M^{(2)}&=t^q\big(-\sigma_0^2T_0\log t+\sigma_0 B+o(1)\big).
 \label{ir:simple}
\end{align}
On the paths constructed in section~\ref{sec:overtaking}, $T_0=0$ and
$B\ne0$. The double pole follows the delayed tree and the simple pole
has order $t^q$.

\subsection{Scale dependence of the two-loop finite part}

Write $\ell=\log(\mu^2/\mu_0^2)$ for a change of subtraction scale at
fixed external momenta. Multiplying the two-loop amplitude by its bare
scale factor $e^{2\ell\epsilon}$ and extracting the finite coefficient
with~\eqref{ir:poles} gives
\begin{equation}
 F_2(\ell)-F_2(0)
 =2\ell\,\sigma F_1(0)+\ell^2\sigma^2\mathcal T ,
 \label{ir:scale}
\end{equation}
where the argument $0$ denotes the reference scale $\mu_0$. For
$T_0\ne0$ and $\ell\ne0$ this gives
\begin{equation}
 F_2(\ell)-F_2(0)=-2\ell\sigma_0^2T_0\,t^q\log t+O(t^q).
 \label{ir:scale-leading}
\end{equation}
This term is nonzero on both original six-point paths and on every
uniform exponential path of section~\ref{sec:one}.
Equation~\eqref{ir:scale} determines the change from the reference value
$F_2(0)$. That value still requires
$[\epsilon^0]\mathcal H^{(2)}+\sigma[\epsilon^1]\mathcal H^{(1)}$
evaluated at $\mu_0$. In the next section we hold $\mu$ fixed and compare
two paths related by an overall momentum dilation. Their difference
will bound the first power of $F_2$ on at least one of them.

\section{Delaying the tree while the one-loop power stays fixed}
\label{sec:overtaking}

On a generic transverse path, section~\ref{sec:one} gives
$F_1/\mathcal T=-\sigma_0\log t+O(1)$. What happens when the leading
tree cofactor vanishes? At six points we can choose a strictly positive
path on which the tree starts at $t^{11}$ while $F_1$ starts at $t^9$.

Write $A$ for negative-helicity leg $1$ and label the $N=n-2$ positive
legs by $j=1,\ldots,N$. For the six-point example take the cyclic order
$(A^-,1^+,2^+,a^-,3^+,4^+)$ of figure~\ref{fig:two-arc-shift}.
In a split-real spin frame, choose the four nodes $z_j=j$ and the
endpoint weights $(1,1,1,-3)$. The path is
\begin{align*}
 \lambda_A&=(1,0),\qquad \lambda_a=(1,t),&
 \lambda_j&=(j,1+jt),\\
 w_1&=w_2=w_3=1,\qquad w_4=-\frac{3+6t}{1+4t},&
 \widetilde\lambda_j&=(w_j,-jt w_j-\delta_{j4}t^2).
\end{align*}
Momentum conservation fixes the two negative-helicity dotted spinors.
The tree cofactor and the signed box sum below give
\begin{align}
 \mathcal T(t)&=
 \frac{t^{11}(4t^2-9t-6)}{6(1+t)(1+2t)(1+3t)}
 =-t^{11}+O(t^{12}),\nonumber\\
 F_1(t)&=-i\pi K_6\,t^9+o(t^9),\qquad
 K_6=72(8\log 2+36\log 3-25\log 5)>0 .
 \label{exc:sixpath}
\end{align}
Here $T_0=0$, so the $t^9\log t$ coefficient vanishes. The finite
constant at the same power gives
$\operatorname{Im}F_1/(\pi t^9)\to-K_6\simeq-349.867586$ and
$F_1/\mathcal T\sim i\pi K_6/t^2$. The integral below shows why the
one-loop constant stays nonzero. By separating the tree roots we can
then prescribe any integer power gap $H\ge1$ for $n\ge6$.

\subsection{Rank-one openings and the one-loop constant}

\begin{theorem}[One-loop power on rank-one openings]\label{exc:integer}
For every $N=m+k$, $m,k\ge2$, a strictly positive ordering with $m$ and
$k$ positive-helicity legs on the two arcs admits an identically vanishing
tree and analytic lifts of exact tree order $q+h(N-2)$ for each integer
$h\ge1$. All have $F_1=B_+t^q+o(t^q)$. In the spin frame below,
\begin{equation}
 \frac{B_+}{i\pi}=-N^2(N-1)^2
 \int_{\prod_{i=1}^{N-1}[0,i]}
 \frac{du_1\cdots du_{N-1}}{u_1+\cdots+u_{N-1}}<0.
 \label{exc:integral}
\end{equation}
\end{theorem}

Take $\lambda_A=(1,0)$, $b=\lambda_a=(1,t)$, $e=(0,1)$ and the following
reflected dotted spin frame. For increasing nodes $z_i$ and $w_i>0$ for
$i<N$, set
\begin{align}
 \lambda_i&=z_i b+e, &v_i&=1+t z_i, &Q_i&=v_i w_i,\nonumber\\
 Q_N&=-\sum_{i<N}Q_i, &w_N&=Q_N/v_N,
 &\widetilde\lambda_i&=(w_i,-t w_i z_i-\delta_{iN} t^{h+1}).
 \label{exc:spinors}
\end{align}
Conservation gives $\widetilde\lambda_a=-\sum_i v_i\widetilde\lambda_i/t$
and $\widetilde\lambda_A=-\widetilde\lambda_a-\sum_i z_i\widetilde\lambda_i$.
Both are analytic. Setting the last perturbation to zero gives the permanent
zero. The integer family uses $z_i=i$, $w_i=1$ for $i<N$.
Equation~\eqref{exc:sixpath} uses $N=4$ and $h=1$. The exact tree
edges are $tQ_iQ_j$, with an extra $t^{h+1}D_i$ on edge $(i,N)$, where
$D_i=Q_i v_N/(z_N-z_i)>0$. Deleting slot $N$ gives the matrix
$t^{h+1}\operatorname{diag}(D_i)-t Q' Q'^{\mathsf T}$, $Q'=(Q_i)_{i<N}$.
The rank-one determinant formula gives
\begin{equation}
 \mathcal{T}=\frac{t^{q+h(N-2)}\prod_{i<N}D_i}{\prod_{i=1}^N v_i^2}
 \left(t^h-\sum_{i<N}Q_i^2/D_i\right).
 \label{exc:lift}
\end{equation}
The parenthesis tends to $M=\sum_i z_i Q_i(0)<0$. Without the perturbation
the deleted matrix has rank one and nullity $N-2$.

\paragraph{Extracting the constant from the boxes.}
Here write $x_i=z_i$, $g_i=Q_i(0)$, $d=N-2$, and first undo the reflection
of the second dotted coordinate. In the affine frame $\lambda_i=(x_i,1)$
at the endpoint the positive dotted spinors start as $(g_i,t g_i x_i)$,
up to tangential motion of their first components and nodes. The negative
dotted limits are $(-M,M),(0,-M)$. All proper nonempty positive subset
sums $g_S=\sum_{i\in S}g_i$ are nonzero because $g_N=-\sum_{i<N}g_i$.
On this one-minority family the unreflected constant satisfies
\begin{equation}
 \frac{B_{\rm u}}{i\pi}=-4\operatorname{sgn}(M) A_N
 \sum_{\varnothing\ne S\subseteq\{1,\dots,N-1\}}
 (-1)^{|S|-1} L_{N,S}^{d}\log|L_{N,S}|,\qquad
 A_j=\frac{M^2\prod_i g_i}{\prod_{i\ne j}(x_i-x_j)},
 \label{exc:subset}
\end{equation}
where $L_{j,S}=\sum_{i\in S}(x_i-x_j)g_i$. We give the reduction since a
nonzero constant at this order is crucial to the two-loop argument.

The channel identity of section~\ref{sec:one} removes the squared
logarithms through order $q$ since $T_0=0$. For a box with massless
corners $A,j$, let the purely positive corner have label set $S$ and the
other corner contain $a$ and $R=P_+\setminus(S\cup\{j\})$, allowing singleton
corners. The two sandwiches in its Gram determinant $D$
start with $-t g_j L_{j,S}$ and $-M L_{j,S}$. With $u=-s_{Aj}/D$ and
$C$ the rational box weight of section~\ref{sec:one}, this gives
\begin{align}
 D&=t M g_j L_{j,S}^2+O(t^2), & u&=(tL_{j,S}^2)^{-1}+O(1),\nonumber\\
 [t^q]C&=(-1)^{|S|-1} A_j L_{j,S}^{d}=:c_{j,S}.
 \label{exc:weights}
\end{align}
The last equality follows from the weighted Cayley sum
$\sum_{\rm trees}\prod_{(ik)} b_i b_k=(\sum b_i)^{|S|-2}\prod b_i$.
Indeed, the two sets of $b_i=(x_i-x_j)g_i$, with $b_a=-M$ on the mixed side,
have sums $L_{j,S}$ and $-L_{j,S}$. Interchanging the negative legs
reverses the signs of $D$ and $c_{j,S}$ and preserves the leading $u$.
The two soft entries of a mirror pair consequently cancel.
For the remaining hard entries put $T'=-Mg_S$ and $P=Mg_R$ at the
endpoint. The phase of a signed dilogarithm divided by $\pi$ is
$\sigma_X\operatorname{sgn}(u)\mathbf{1}_{uX<0}\log(1-uX)$.
The active hard logarithm is $-\log t+\log|X(0)|-2\log|L_{j,S}|+o(1)$.
The channel identity cancels the first two pieces, and the mirror
coefficient of $\log|L_{j,S}|$ is
$-2c_{j,S}(\operatorname{sgn}P-\operatorname{sgn}T')$.
For $R\ne\varnothing$ it vanishes unless $j=N$, in which case the
sign difference is $2\operatorname{sgn}M$. For $R=\varnothing$ set
$\operatorname{sgn}P=0$ and use $L_{j,S}=M$ and $\sum_j A_j=0$
(Lagrange interpolation). This supplies the full-set term in
\eqref{exc:subset}. Boxes with $L_{j,S}=0$ have zero weight at this
order, and $L^d\log|L|$ extends to zero there.

For completeness the stars contribute no phase at this order.
In each star $u=O(t)$ and the Gram relation implies
$\sum_X\sigma_X\log(1-uX)=0$. Its phase therefore equals
$-\pi\sum_{X>0}\sigma_X\log(1-uX)$. The hard sign for a side containing
$A$ depends only on whether it also contains $N$. If $N$ is in a massive
corner the sign is common to all entries. If $N$ is a massless corner,
the sign for each entry is independent of the remaining $N-2$ corner
assignments. Their alternating sum vanishes through relative degree
$N-3$ by the forest degree bound of section~\ref{sec:one}.
The real parts of opposite hard dilogarithms in a mirror pair differ,
at this order, by a constant times $\operatorname{sgn}X(0)$. The weighted
sum vanishes by the same hard-channel identities. For the real star
remainder use parity. In the affine frame exchange
the negative labels, apply $(\lambda^1,\lambda^2)\mapsto
(\lambda^1,\lambda^2-t\lambda^1)$ and
$(\tilde\lambda^{\dot1},\tilde\lambda^{\dot2})\mapsto
(\tilde\lambda^{\dot1}+k\tilde\lambda^{\dot2},-\tilde\lambda^{\dot2})$
with $k=-\tilde\lambda_A^{\dot1}/\tilde\lambda_A^{\dot2}\to1$, and
restore the affine positive frames. This fixes the endpoint and negates
the normal tangent $(\langle Aa\rangle,(\tilde\lambda_i^{\dot2})_{i<N})'$.
The summed real star kernel is analytic in these normal coordinates and
vanishes below degree $q$ throughout the nearby regular tangential data
by section~\ref{sec:one}. Its degree-$q$ term therefore picks up $(-1)^q$.
All invariants reverse, every $uX$ is fixed, and the square brackets in
the rational weight give $(-1)^{n-4}$. The opposite parities force this
term to zero. This argument also explains why tangential motion and the
higher lift in \eqref{exc:spinors} preserve the coefficient. Only the endpoint and normal tangent enter the analytic
star part, and the nonstar
part uses the leading weights, hard endpoints and first soft derivatives.

For $j=N$, set $r_i=(x_N-x_i)g_i>0$. Then $L_{N,S}=-\sum_{i\in S}r_i$.
Writing $f(y)=y^d\log y$, $f(0)=0$, the signed sum in \eqref{exc:subset}
is its $(d+1)$-fold forward difference, with positive increments $r_i$.
Since $f^{(d+1)}(y)=d!/y$, it equals
$d!\int_{\prod[0,r_i]}(\sum u_i)^{-1}\prod du_i$. This follows first at
positive base point and then by monotone convergence, since $N-1>1$ ensures
integrability at the origin. For integer nodes the increments are
$1,\dots,N-1$ and $A_N=(-1)^N M^2/d!$. Dotted reflection conjugates
the causal dilogarithms and multiplies the rational weight by
$(-1)^{n-4}$. Thus $B_+=(-1)^{n-4}\overline{B_{\rm u}}$, which gives
\eqref{exc:integral} since $M=-N(N-1)/2$. For $N=4$ the three increments
are $1,2,3$. The forward difference reduces to
$6^2\log 6-5^2\log 5-4^2\log 4+2^2\log 2$, yielding $K_6$ in
\eqref{exc:sixpath}. The same calculation gives
$\operatorname{Im} B_+<0$ for any increasing nodes with the single
negative weight at the largest node.

\paragraph{Strict positivity of the incidence twistors.}
The following test works both for the integer example and for the two
orderings used below. Choose $m$ majority nodes as a left set $I$ and
order $(A,I,a,I^c)$ with both arcs increasing. In
$Z_i=(\lambda_i,\sum_{j<i}\langle ji\rangle\widetilde\lambda_j)$,
negate the columns on $I$ and the column $a$. Write $U_p,V_p,W_p$ for
the cumulative sums of $g_i,g_i x_i,g_i x_i^2$ over the first $p$
positive nodes in this order and $H_0=-M>0$. After subtracting
$H_0\lambda^2$ from the third twistor row the endpoint private columns,
before signs, have coordinates $(x,1,-U_p x+V_p,H_0)$ on the left and
$(x,1,-U_p x+V_p,0)$ on the right for $p$ predecessors. The unsigned
seams are $(1,0,0,0),(1,0,-U_m,0)$. Left and right secant slopes $s_L,s_R$
of the third coordinate obey $-U_m<s_L<0$ and $s_R<-U_m$. All slopes
strictly decrease along either arc. Subtracting columns now gives the
following mixed minors divided by $H_0$. Here $L,R$ denote private
columns, $\Delta$ a positive pair separation, and
$V_3=(x_j-x_i)(x_l-x_j)(s_{ij}-s_{jl})>0$ on any private triple.
\[
\begin{array}{c|c@{\qquad}c|c}
 LLLR,\ LRRR& V_3 & LLRR & \Delta_L\Delta_R(s_L-s_R)\\
 ALLR & -\Delta_L s_L & ALRR &-\Delta_R s_R\\
 LLaR & \Delta_L(U_m+s_L)&LaRR&\Delta_R(-U_m-s_R)\\
 ALaR & U_m &&
\end{array}
\]
Every pure minor (all four columns on one closed arc) vanishes at zero.
To obtain its linear term, subtract the affine part of row four on that
arc and divide by $t$. For fixed affine nodes and first charges, if the
unreflected second dotted components are $t g_ix_i+t^2 e_i+O(t^3)$,
the left subtraction is $[H_0+t(W_N-\sum e_i)]\lambda^2$, the right
one $t(W_N\lambda^2+H_0\lambda^1)$. Direct incidence summation gives
private columns $(x,1,-U_p x+V_p,V_p x-W_p)$. A segment displacement
is its positive length times $D_p=(1,0,-U_p,V_p)$. The left seams are
$D_0,D_m$ and the right seams $D_m,D_N$ before their signs. For moving
affine nodes and charges the linear coefficient is unchanged because the pure minor at zero
vanishes identically in all tangential parameters.

Set $d_p=(1,-U_p,V_p)$. Taking successive private-column differences
and expanding the row of ones gives a positive length sum of
$-\det(d_i,d_j,d_l)$ with $i<j<l$, with an initial or terminal seam supplying
the first or last direction. On the right the terminal seam changes the
sign, from moving $A$ past the other three columns. When $l<N$ the
negative determinant equals $\sum_{i<s\le j<u\le l}g_sg_u(x_u-x_s)>0$
with positions read on that arc. For $l=N$ on the right the required sign is
$\det(d_i,d_j,d_N)>0$. Set $\bar x=(V_j-V_i)/(U_j-U_i)$. The determinant
is $(U_j-U_i)[U_i(x_N-\bar x)+\sum_{i<s<N}g_s(x_N-x_s)]>0$.
This proves positivity of every mixed constant and every pure linear
coefficient, hence strict positivity for the finitely many minors on
a punctured interval. This finishes the proof of Theorem~\ref{exc:integer}.

\subsection{Tuning the tree power with a two-leg deformation}

\begin{theorem}[Arbitrary power gaps at fixed scale]\label{exc:overtake}
At fixed subtraction scale, for each $n\ge6$ and integer $H\ge1$ there
exist strictly positive real analytic paths with
\begin{equation}
 \operatorname{ord}_t\mathcal{T}=q+H,\qquad
 \operatorname{ord}_t F_1=q,\qquad \operatorname{ord}_t F_2\le q.
 \label{exc:orders}
\end{equation}
In each of two backgrounds a displacement from a tree-zero path changes
only one negative- and one positive-helicity momentum. The same
prescribed mixed two-particle invariant $S>0$ is held fixed in both.
For each prescribed nonzero leading displacement, one can choose a path
in each background such that at least one realizes all three orders.
\end{theorem}

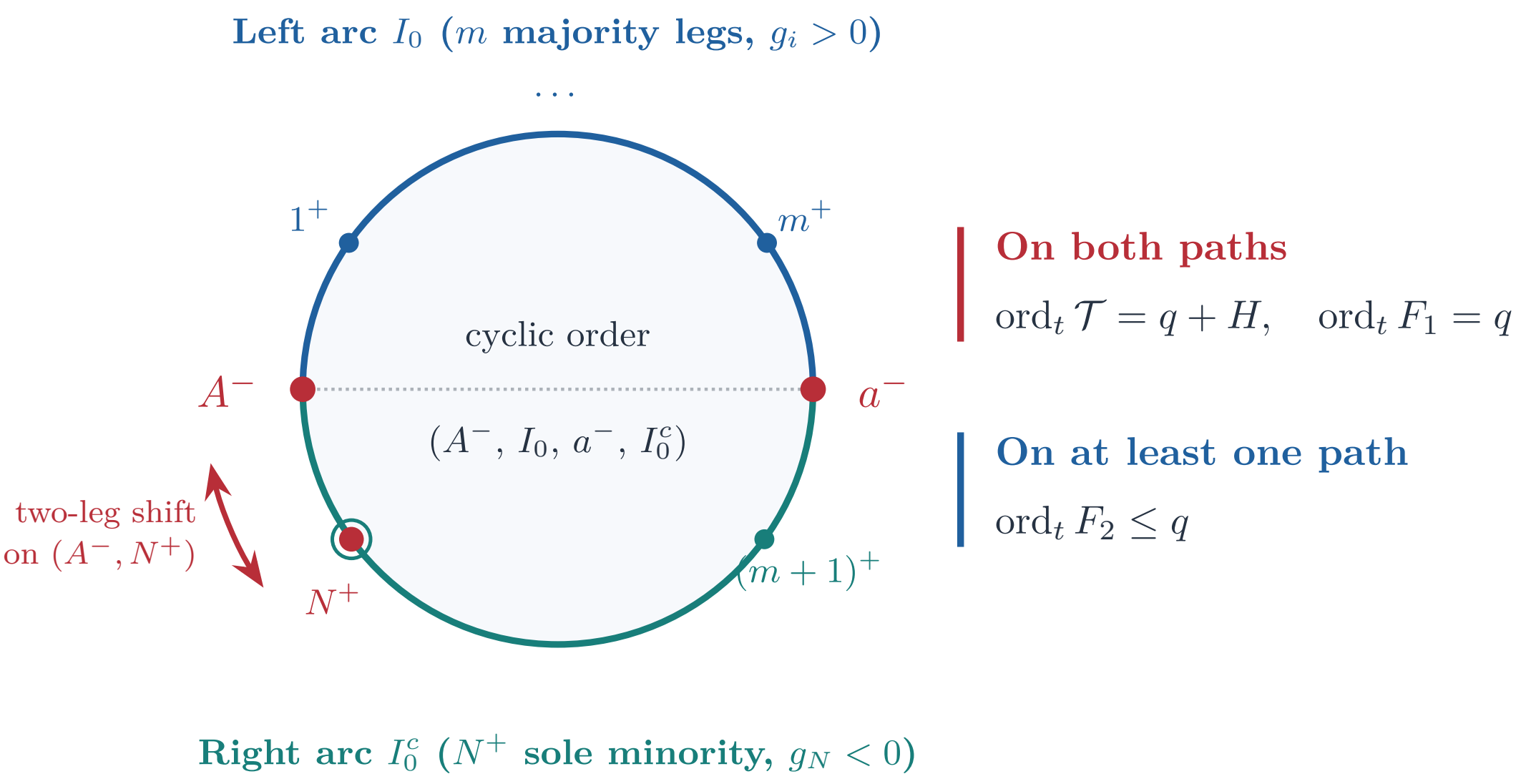
\begin{figure}[t]
\centering
\resizebox{0.96\textwidth}{!}{%
\begin{tikzpicture}[>=Stealth, font=\small]

\def\R{2.25}

\fill[ampDiskFill] (0,0) circle (\R);
\draw[draw=ampMuted!55, densely dotted, line width=0.8pt] (-\R,0) -- (\R,0);

\draw[draw=ampBlue, line width=1.65pt] (180:\R) arc[start angle=180, end angle=0, radius=\R];
\draw[draw=ampTeal, line width=1.65pt] (0:\R) arc[start angle=0, end angle=-180, radius=\R];

\node[text=ampSlate, font=\footnotesize, align=center] at (0, 0.46)
  {cyclic order};
\node[text=ampSlate, font=\footnotesize, align=center] at (0, -0.46)
  {$(A^-,\, I_0,\, a^-,\, I_0^c)$};

\fill[ampRed] (180:\R) circle (3.2pt);
\fill[ampRed] (0:\R) circle (3.2pt);
\node[text=ampRed, font=\bfseries, anchor=east] at (180:\R+0.26) {$A^-$};
\node[text=ampRed, font=\bfseries, anchor=west] at (0:\R+0.26) {$a^-$};

\foreach \ang/\lab in {145/{$1^+$}, 35/{$m^+$}}{
  \fill[ampBlue] (\ang:\R) circle (2.5pt);
  \node[text=ampBlue, font=\footnotesize] at (\ang:\R+0.42) {\lab};
}
\node[text=ampBlue, font=\footnotesize] at (90:\R+0.34) {$\cdots$};
\node[text=ampBlue, font=\bfseries\footnotesize] at (90:\R+0.88)
  {Left arc $I_0$ ($m$ majority legs, $g_i>0$)};

\fill[ampTeal] (-36:\R) circle (2.5pt);
\node[text=ampTeal, font=\footnotesize] at (-36:\R+0.48) {$(m+1)^+$};
\fill[ampRed] (-144:\R) circle (3.1pt);
\draw[draw=ampTeal, line width=0.9pt] (-144:\R) circle (4.8pt);
\node[text=ampRed, font=\bfseries\footnotesize] at (-137:\R+0.46) {$N^+$};

\node[text=ampTeal, font=\bfseries\footnotesize] at (-90:\R+0.96)
  {Right arc $I_0^c$ ($N^+$ sole minority, $g_N<0$)};

\draw[draw=ampRed, line width=1.35pt, <->]
  (192:\R+0.88) arc[start angle=192, end angle=214, radius=\R+0.88];
\node[text=ampRed, font=\scriptsize, align=right, anchor=east] at (203:\R+1.06)
  {two-leg shift\\on $(A^-, N^+)$};

\begin{scope}[shift={(3.55, 0)}]
  \draw[draw=ampRed, line width=1.7pt] (0, 1.43) -- (0, 0.42);
  \node[anchor=west, text=ampSlate, align=left] at (0.17, 0.93) {%
    \textbf{\color{ampRed}On both paths}\\[4pt]
    $\operatorname{ord}_t\mathcal{T}=q+H,\quad \operatorname{ord}_t F_1=q$
  };
  \draw[draw=ampBlue, line width=1.7pt] (0, -0.38) -- (0, -1.39);
  \node[anchor=west, text=ampSlate, align=left] at (0.17, -0.88) {%
    \textbf{\color{ampBlue}On at least one path}\\[4pt]
    $\operatorname{ord}_t F_2\le q$
  };
\end{scope}

\path ([xshift=-4pt,yshift=-4pt]current bounding box.south west)
      ([xshift=4pt,yshift=4pt]current bounding box.north east);

\end{tikzpicture}%
}
\caption{Leading powers on two paths at fixed subtraction scale $\mu$ and
mixed invariant $S>0$ (Theorem~\ref{exc:overtake}). The diagram shows the
ordering with $I_0=\{1,\ldots,m\}$ and $N=m+2$. The second uses
$I_1=\{2,\ldots,m+1\}$. Within either background the adjacent legs
$A^-,N^+$ exchange a null momentum and the other $n-2$ momenta are fixed
at each $t$. The pair $(a^-,(m+j)^+)$ is normalized to $S$ in
background $j=0,1$.}
\label{fig:two-arc-shift}
\end{figure}

A small change of background from $d_1=d_0$ to $d_1>d_0$ separates the
$N-2$ coincident hard tree zeros into simple branches $z=Z_j(t)$.
The on-shell shift of the adjacent legs $(A^-,N^+)$ in
figure~\ref{fig:two-arc-shift} holds the other $n-2$ momenta fixed at
each $t$ and follows one such branch through order $H-1$ before
departing at order $H$. Even for $H=1$ this displacement is tangential
at the endpoint, so the locally uniform one-loop expansion below
retains the branch coefficient $B_*$.

Keep the angles of \eqref{exc:spinors}. Take $d_0=2$ and $d_1>d_0$
sufficiently close to $d_0$. For $i<N$, set
$z_i=d_0^{i-1}$, $w_i=(d_0 d_1)^{-(i-1)}$ and $f_i=f(z_i)$ with
$f(z)=z^{\log_{d_0}d_1}$. The last node $z=z_N$ varies near
$z_*=d_0^{N-1}$, while its ordinate $f_N=y$ will be fixed to $z_*$.
Set $Q_i=v_iw_i$ as before, keep $Q_N=-W=-\sum_{i<N}Q_i$, and use
$\widetilde\lambda_i=(w_i,-t w_i f_i)$. Then
$\widetilde\lambda_a=(0,\sum_i Q_i f_i)$ is independent of $z$.
The exact tree is $t^q\det L_*(y)/\prod_i v_i^2$, where $L_*$ is the
slot-$N$ deleted Laplacian with edges $Q_iQ_j(f_i-f_j)/(z_i-z_j)$.

Write $p_i=Q_i(z-z_i)>0$ for $i<N$. The rational function
$\sum_{i<N} p_i/(r-z_i)$ decreases from $+\infty$ to $-\infty$ on each
$(z_j,z_{j+1})$, $j\le N-2$. Denote its simple root by $r_j$ and set
\[
 \bar f=W^{-1}\sum_{i<N}Q_i f_i,\qquad
 Y_j=\bar f+W^{-1}\sum_{i<N}p_i\frac{f(r_j)-f_i}{r_j-z_i}.
\]
Strictly increasing, strictly convex $f$ gives
$\bar f<Y_1<\cdots<Y_{N-2}$. The determinant factorizes as
\begin{equation}
 \det L_*(y)=\left(\prod_{i<N}\frac{Q_iQ_N}{z-z_i}\right)
 (y-\bar f)\prod_{j=1}^{N-2}(y-Y_j).
 \label{exc:factor}
\end{equation}
In fact $\sum_iQ_i/(r_j-z_i)=0$ and at $y=Y_j$ also
$\sum_i Q_i f_i/(r_j-z_i)=0$. Subtracting the components across each
edge shows that $1/(r_j-z_i)$ is a second null vector of the full
Laplacian. At $y=\bar f$ the additional null vector has components $z_i$. Subtracting a constant
sets the last component to zero. Finally the coefficient of $y$ in
$L_*$ is diagonal with entries $Q_iQ_N/(z-z_i)$, proving
\eqref{exc:factor}. The root $\bar f$ makes $a$ soft and is separated
from the hard roots in this construction.

For $d_1=d_0$, $f$ is linear and every $Y_j=z$ identically, while the
$r_j$ remain simple. Hence $\partial_zY_j=1$ there. For fixed
sufficiently small $d_1-d_0>0$, each equation $Y_j(z,t)=z_*$ has an
analytic solution $Z_j(t)$ near $z_*$. They are distinct and simple by
the strict ordering of the $Y_j$ and persistence of $\partial_zY_j>0$.
Clearing the denominators by multiplying row $i<N$ of $L_*(z_*)$ by
$z-z_i$ gives a polynomial determinant of degree at most $N-1$. Its
top coefficient vanishes because the majority-graph Laplacian kills
the constants. The $N-2$ simple roots just found exhaust its degree.
Along
\begin{equation}
           z=Z_j(t)+c t^H,\qquad c\in\mathbb R\setminus\{0\},
 \label{exc:tune}
\end{equation}
the slope of the cleared determinant has a nonzero limit. Thus the
order is exactly $q+H$. In the affine minority frame
$\lambda'_N=(z/(1+tz),1)$ and $\widetilde\lambda'_N=Q_N(1,-t z_*)$.
A displacement of $z/(1+tz)$ by $\delta$ adds exactly
$\delta\lambda_A(\widetilde\lambda'_N)^{\mathsf T}$ to $p_N$ and
subtracts it from $p_A$. All $n-2$ spectator momenta are independent
of $z$ at each ordinary $t$.\footnote{Feng, Hu and Zhang used a two-leg shift of this form to
approach a factorization pole and track the resulting collisions of
punctures in the scattering equations~\cite{FengHuZhang}.}

At $d_1=d_0$ and $z=z_*$, these data reduce to the permanent reflected
rank-one path, so the strict minor inequalities persist for small
$d_1-d_0$ on each root; since the pure constants continue to vanish
identically, their linear coefficients still decide the signs, and the
grouped one-loop coefficients vary continuously. In detail the hard subset sums are nonzero by the
one-minority sign. For $j<N$, $N\notin S$, any term of $L_{j,S}$ with
$i<j$ dominates the sum of possible positive terms with $i>j$ by the
geometric hierarchy $g_i=4^{-(i-1)}$, $z_i=2^{i-1}$. When $N\in S$ its
term dominates all positive terms, and for $j=N$ every term has one
sign. The first pure soft invariants, proportional to
$\sum_{i<j\in S}g_i g_j(z_i-z_j)^2$, have one sign when $N\notin S$.
If $N\in S$, for each $i<N$ its minority contribution dominates the
sum over ordinary $j>i$ since $|g_N|>\sum_{i<j<N}g_j$ and $z_N>z_j$.
This gives a locally uniform expansion of the nonstar boxes. The grouped star
kernel is analytic. At $d_1=d_0$ the extreme-node integral gives
$\operatorname{Im} B_+<0$, so for sufficiently small fixed $d_1-d_0>0$
the finite $t^q$ coefficient $B_*$ on each tuned branch still has
negative imaginary part. Its $\log t$ coefficient vanishes by the
cofactor test. On substituting $z=Z_j(t)+ct^H$ in this uniform
expansion, the coefficient of $\log t$ changes by $O(t^H)$ and the
constant coefficient tends to its value on the same root. Since
$t^H\log t\to0$, the limit of $t^{-q}F_1$ is $B_*$ on the displaced path.

\paragraph{The two-loop power at fixed scale.}
Under a common positive dilation of all momenta by $e^x$ each spinor
scales by $e^{x/2}$. Two-loop homogeneity gives
$M^{(2)}(e^x p;\epsilon)=e^{(6-4\epsilon)x}M^{(2)}(p;\epsilon)$ at fixed
$\mu$. At two loops four-dimensional maximal supergravity has no allowed
on-shell ultraviolet counterterm~\cite{EFK}. Expanding the two infrared poles~\eqref{ir:poles} of section~\ref{sec:ir}
gives the exact relation
\begin{equation}
 G_2(x):=e^{-6x}F_2(e^x p)=F_2(p)-4x\sigma F_1+4x^2\sigma^2\mathcal T.
 \label{exc:energy}
\end{equation}
Here $\sigma=\sum_{i<j}s_{ij}\log((-s_{ij}-i0)/\mu^2)$ and
$\sigma(0)=\sigma_0=-i\pi\sum_{i\in P_+}|s_{Ai}(0)|\ne0$.

For an explicit pair of backgrounds take $m=N-2$, $k=2$. In the
positive orders above choose left subsets $I_0=\{1,\ldots,m\}$,
$I_1=\{2,\ldots,m+1\}$, and read backwards starting at $a$. Reversal
preserves the signs of the four-minors. The associated change of incidence
convention negates both lower twistor rows and has determinant $+1$.
The first two slots are now $a,m+j$ in background $j=0,1$. Their hard
invariant $s_j(t)>0$ is proportional to the majority weight $w_{m+j}$,
with the same factor $-\sum_iQ_i f_i$ for both, independent of $z$.
Dilate by $e^{x_j(t)}=\sqrt{S/s_j(t)}$ and apply a proper split Lorentz
transformation to match the two momenta to one fixed pair. Use the two independent
angle spinors and the two independent dotted spinors as bases. Since both
pairs have invariant $S$, reciprocal spin-frame rescalings match the
two basis determinants separately. This gives determinant-one changes of basis with analytic unit limits.
The transformations depend only on the spectators. Thus the two-leg
definition of the displacement still holds within each background, the
mixed invariant agrees pointwise, and $x_1-x_0=\tfrac12\log(d_0 d_1)\ne0$.
Gravity is invariant under the simultaneous relabeling of its external
momenta and helicities. In transported spin frames, \eqref{exc:energy}
therefore gives on \eqref{exc:tune}
\begin{equation}
 G_2(x_1)-G_2(x_0)=-4(x_1-x_0)\sigma_0 B_*\,t^q+o(t^q).
 \label{exc:difference}
\end{equation}
Since $\sigma_0$ is purely imaginary and $\operatorname{Im}B_*<0$,
the coefficient has nonzero real part. Consequently $F_2$ has order
at most $q$ on at least one of the two paths. Finite analytic
spin-frame and energy units preserve that order. At one loop the
energy shift adds a multiple of $\sigma\mathcal T$ to the normalized
$F_1$ and therefore also preserves its order $q$.

\paragraph{Power counting and the choice of ordinary analytic rays.}
On an ordinary analytic ray, a Mellin transform in $t$ gives the
expansion in powers of $t$ and $\log t$ used to define $\operatorname{ord}_t$.
In the dimensionally regulated Feynman parameter integral, resolve
the product of $t$, the faces and the Symanzik polynomials
on the compact simplex and retain the $-i0$ phase on each sign orthant.
The local Mellin integrals have factors
$y_i^{a_i s+b_i+c_i\epsilon-1}$ times smooth units with rational
$b_i,c_i$ and integers $a_i\ge0$. Taylor subtraction gives poles at
$a_i s+b_i+c_i\epsilon=-j$, $j\ge0$. The Laurent coefficient
and Mellin residues give discrete rational powers with logarithmic
polynomials. This argument uses the ordinary boundary value for $t>0$.
To obtain ordinary rays for any fixed $H$ and $c\ne0$, take local
on-shell coordinates $(z,\eta)$, with $z$ along the two-leg shift and
$\eta$ labelling its lines. The shift vector is nonzero and
$z\mapsto z/(1+tz)$ is locally invertible. Perturb $\eta$ by $t^K v$
with an integer $K>q+H$. The normalized tree is analytic, so its
simple root persists as $Z_j(t;v)$ with the required jets. Allow a
further displacement along the same line,
\[
                 z=Z_j(t;v)+ct^H+\alpha t^K .
\]
At fixed small $t>0$, varying $(v,\alpha)$ covers an open on-shell
neighborhood while $c$ stays fixed. Choose $(v,\alpha)$ there to avoid
the finitely many additional proper algebraic singular loci. Their
defining functions are analytic along the path, so they remain
nonzero on a sufficiently small punctured interval. The two pair
normalizations preserve ordinary kinematics by permutation, proper
Lorentz transformation and positive dilation. The high-order perturbations
preserve the strict minor signs, the two distinct limiting energies and
the leading term of \eqref{exc:difference}. A term shared by the two
normalized amplitudes can start below order $q$ and cancel in their
difference. This leaves the possibility of a still earlier two-loop power.

\section{The negative-helicity pair cut at two loops}
\label{sec:cut}

Cut the two-loop amplitude in the channel $s=(p_1+p_a)^2$ and place
the loop on the side carrying the negative-helicity pair. The two
external gravitons $1^-,a^-$ saturate the Grassmann degree sixteen of
the four-point superamplitude. Both cut gravitons therefore have
positive helicity on this side and negative helicity on the $n$-point
tree across the cut.

In the forward and backward directions that tree tends to the same
external tree amplitude $\mathcal T$. We subtract this endpoint value,
sew the difference to the four-point finite part $F_4^{(1)}$, and
integrate over the four-dimensional hard directions. The resulting
angular moment $\mathcal C_4$ can be evaluated on a timelike Lorentzian
sphere and continued to the split-signature limit. On the two original
six-point paths the loop insertion carries an extra factor of $s=O(t)$
relative to the four-point tree, and the moment starts at $t^{10}\log^2 t$.
The jointly subtracted pole in section~\ref{cut:soft} retains the tree
angular moment and starts at $t^9$ on these paths.

\subsection{Integration over the hard directions}

\begin{figure}[htbp]
\centering
\resizebox{\textwidth}{!}{%
\begin{tikzpicture}[>=Stealth, font=\small]

\begin{scope}[shift={(0,0)}]
  \node[font=\bfseries, text=ampSlate, align=center, anchor=south] at (3.20, 2.84)
    {(a) Four-point one-loop contribution\\to the pair cut};

  \draw[neg leg] (0.05,2.05) -- (0.85,1.65);
  \draw[neg leg] (0.05,0.05) -- (0.85,0.45);
  \node[text=ampRed, anchor=east] at (0.02,2.08) {$1^-$};
  \node[text=ampRed, anchor=east] at (0.02,0.02) {$a^-$};

  \fill[ampSoftFill] (0.85,0.45) rectangle (2.05,1.65);
  \draw[hard line] (0.85,0.45) rectangle (2.05,1.65);
  \node[text=ampBlue, font=\footnotesize] at (1.45,1.05) {$F_4^{(1)}$};

  \draw[hard line, mid arrow] (2.05,1.65) -- (5.05,1.65);
  \draw[hard line, mid arrow] (2.05,0.45) -- (5.05,0.45);
  \node[text=ampSlate, font=\footnotesize, anchor=south] at (2.48,1.72) {$Q_1$};
  \node[text=ampSlate, font=\footnotesize, anchor=north] at (2.48,0.38) {$Q_2$};

  \draw[cut line] (3.20,-0.15) -- (3.20,2.25);
  \node[text=ampRed, font=\scriptsize, anchor=south] at (3.20,2.28) {$s=(p_1+p_a)^2$};

  \node[text=ampBlue, font=\scriptsize] at (2.92,1.43) {$+$};
  \node[text=ampRed,  font=\scriptsize] at (3.48,1.43) {$-$};
  \node[text=ampBlue, font=\scriptsize] at (2.92,0.67) {$+$};
  \node[text=ampRed,  font=\scriptsize] at (3.48,0.67) {$-$};

  \draw[pos leg] (5.05,1.05) -- (6.20,1.95);
  \draw[pos leg] (5.05,1.05) -- (6.30,1.40);
  \draw[pos leg] (5.05,1.05) -- (6.30,0.70);
  \draw[pos leg] (5.05,1.05) -- (6.20,0.15);
  \node[hard blob, minimum size=42pt] at (5.05,1.05) {};
  \node[text=ampSlate, font=\footnotesize] at (5.05,1.05) {$T_n'\!-\!\mathcal{T}$};
  \node[text=ampSlate, font=\footnotesize, anchor=west] at (6.32,1.05) {$P_+$};

  \node[text=ampSlate, font=\footnotesize] at (3.20,-0.78)
    {$[t^{n+4}\log^2 t]\,\mathcal{C}_4=4i\pi\kappa^2 S_c$};
\end{scope}

\begin{scope}[shift={(8.45,0)}]
  \node[font=\bfseries, text=ampSlate, anchor=south] at (3.20, 2.84)
    {(b) Three-particle soft cut $E^{[-1]}$};

  \draw[neg leg] (0.05,2.05) -- (1.45,1.05);
  \draw[neg leg] (0.05,0.05) -- (1.45,1.05);
  \node[text=ampRed, anchor=east] at (0.02,2.08) {$1^-$};
  \node[text=ampRed, anchor=east] at (0.02,0.02) {$a^-$};

  \draw[hard line, mid arrow] (1.45,1.52) -- (5.05,1.52);
  \draw[hard line, mid arrow] (1.45,0.58) -- (5.05,0.58);
  \draw[soft grav] (1.45,1.05) -- (5.05,1.05);

  \node[hard blob, minimum size=36pt] at (1.45,1.05) {};
  \node[text=ampSlate, font=\footnotesize] at (1.45,1.05) {$T_5$};

  \node[text=ampSlate, font=\footnotesize, anchor=south] at (2.48,1.58) {$Q_1$};
  \node[text=ampSlate, font=\footnotesize, anchor=north] at (2.48,0.52) {$Q_2$};
  \node[text=ampTeal, font=\scriptsize, anchor=south] at (2.60,1.12) {soft $k$};

  \draw[cut line] (3.20,-0.15) -- (3.20,2.25);
  \node[text=ampRed, font=\scriptsize, anchor=south] at (3.20,2.28) {$3$-particle cut};

  \draw[pos leg] (5.05,1.05) -- (6.20,1.95);
  \draw[pos leg] (5.05,1.05) -- (6.30,1.40);
  \draw[pos leg] (5.05,1.05) -- (6.30,0.70);
  \draw[pos leg] (5.05,1.05) -- (6.20,0.15);
  \node[hard blob, minimum size=42pt] at (5.05,1.05) {};
  \node[text=ampSlate, font=\footnotesize] at (5.05,1.05) {$T_{n+1}$};
  \node[text=ampSlate, font=\footnotesize, anchor=west] at (6.32,1.05) {$P_+$};

  \node[text=ampSlate, font=\footnotesize] at (3.20,-0.78)
    {$8\,\dfrac{dE^{[-1]}}{d\nu} = T_4 T_n'(\sigma_E-\sigma_L-\sigma_R-2i\pi s)$};
\end{scope}

\path ([xshift=-4pt,yshift=-4pt]current bounding box.south west)
      ([xshift=4pt,yshift=4pt]current bounding box.north east);

\end{tikzpicture}%
}
\caption{Two-loop cuts in the channel $s=(p_1+p_a)^2$. \textbf{(a)}
The square denotes the full four-point finite subamplitude, summed over
six ordered boxes. Its sewing to the endpoint-subtracted tree defines
$\mathcal C_4$ in~\eqref{cut:periods}. The displayed double-logarithmic
coefficient gives the leading term when $S_c\ne0$.
\textbf{(b)} The three-particle cut sews a five-point tree to an
$(n+1)$-point tree. As $k=\rho k_0$ becomes soft, the two Weinberg
factors multiply the hard trees $T_4 T'_n$. Integrating the soft
direction and extracting the radial $1/\epsilon$ pole gives the
density in~\eqref{cut:softdensity} at fixed hard directions.}
\label{fig:two-loop-cuts}
\end{figure}
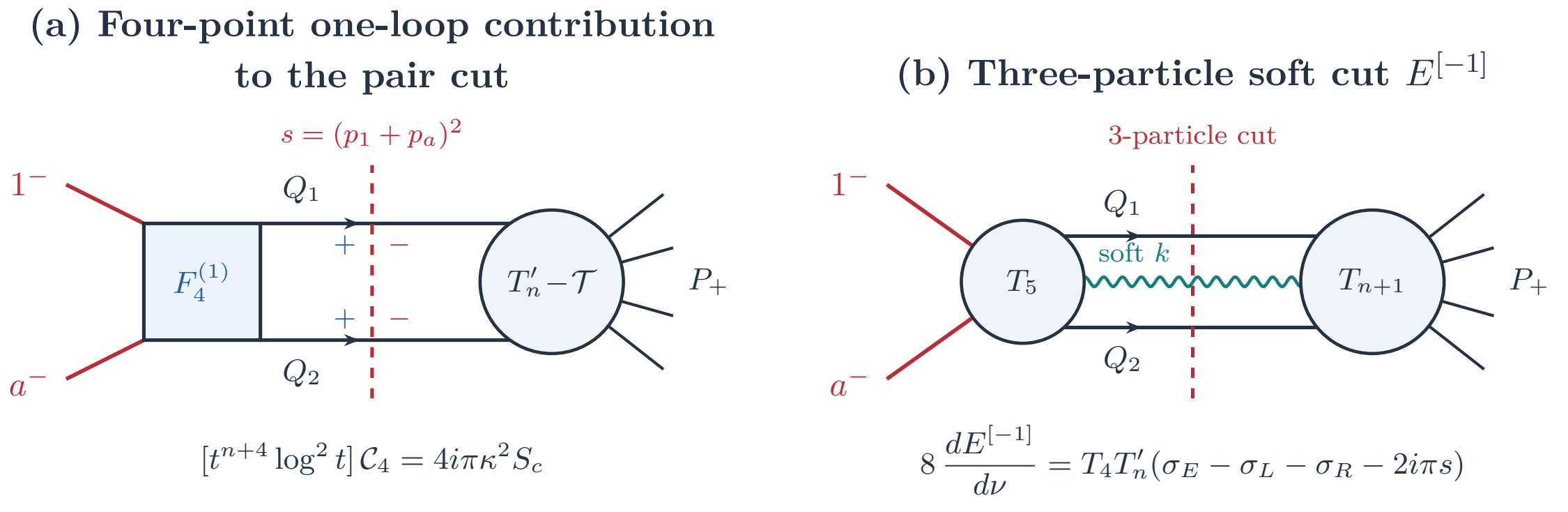

Start at a generic Lorentzian point with $s>0$. Orient $p_1,p_a$ and the
two null momenta $Q_1,Q_2$ in $Q_1+Q_2=p_1+p_a$ towards the future,
and the remaining external momenta towards the past. Set
$x=(1-\cos\vartheta)/2$ in the pair rest frame and normalize the measure to
$d\nu=dx\,d\phi/(2\pi)$. The four-point invariants are $s,-sx,-s(1-x)$.
In a cut-spinor frame of unit determinant the four-point tree is
$T_4=-s/[x(1-x)]$. Denote the tree with external negative legs replaced by
$Q_1,Q_2$ by $T'_n$. Its forward and backward limits both give
$\mathcal T$, allowing for exchange of the two gravitons. Define
\begin{equation}
 \mathcal B_{\rm sph}=\int d\nu\,T_4(T'_n-\mathcal T),\qquad
 \mathcal C_4=\int d\nu\,F^{(1)}_4(T'_n-\mathcal T).
 \label{cut:periods}
\end{equation}
Figure~\ref{fig:two-loop-cuts} shows the two-particle sewing and the
associated soft three-particle cut. Both integrals in~\eqref{cut:periods}
converge. The azimuthal mean of $T'_n-\mathcal T$ vanishes linearly in
$x$ or $1-x$ at each endpoint, while an ordinary collinear pole grows
at most as the inverse angular distance in two real dimensions.

For positive labels introduce the following invariants and angle ratios,
\begin{align}
 b_j=s_{1j},\quad c_j=s_{aj},\quad U_j=b_j+c_j,\quad
 &G_{ij}=U_iU_j-s\,s_{ij},\nonumber\\
 X_j=\frac{\langle1j\rangle}{\langle aj\rangle},\quad
 \theta_j=\frac{b_j}{U_j},\quad
 &\zeta_{ji}=\frac{X_j}{X_j-X_i},\qquad \zeta_{ij}=1-\zeta_{ji}.
 \label{cut:angles}
\end{align}
In the rest frame the negative angles on the right can be taken as
$\lambda_1+z\lambda_a$ and $(1-x)\lambda_a-x\lambda_1/z$.
On the azimuthal circle $|z|^2=x/(1-x)$ in normalized spinor frames.
A positive vertex $j$ therefore has tree weight proportional to
$(z+X_j)(z-xX_j/(1-x))/z$. In a tree monomial its exponent is
$\deg_j-2$. The sum of these exponents is $-2$, so $T'_n$ vanishes
quadratically at zero and as $z^{-2}$ at infinity. A leaf alone supplies
a pole at $z=-X_j$. Let $A_j(x)$ be the residue of $T'_n\,dz/z$ there.
The other pole has residue $A_j(1-x)$ by exchange of the two ports.
The relation $|X_j|^2=b_j/c_j$ gives the step functions in the mean,
\begin{equation}
 \overline{T'_n}(x)=\sum_j
 \left[\mathbf 1_{x>\theta_j} A_j(x)
       +\mathbf 1_{x<1-\theta_j} A_j(1-x)\right],\qquad
 \sum_j A_j(1)=\mathcal T .
 \label{cut:azimuth}
\end{equation}

After taking the residue at $j$, the remaining vertex weights are affine
in $x$. A second leaf $i$ gives a pole at $\zeta_{ji}$. The sum of
the exponents on the remaining vertices is $-1$, so there is no
polynomial part. In terms of the half-soft function of
section~\ref{sec:one}, the result is
\begin{equation}
 A_j(x)=\sum_{i\ne j}\frac{B_{ij}}{x-\zeta_{ji}},\qquad
 B_{ij}=-\frac{\langle1a\rangle^4}{\langle ij\rangle^2}\,
             G_{ij}\,h(i,P_+\setminus\{i,j\},j)=B_{ji}.
 \label{cut:doubleleaf}
\end{equation}
Indeed, at the second pole the cut angles are proportional to those of
$j$ and $i$. Deleting both leaves leaves the indicated half-soft tree.
The independent sums over their attachments multiply to
$\langle i|p_1+p_a|j]\langle j|p_1+p_a|i]=G_{ij}$, and the two
weight derivatives have product
$-\langle1a\rangle^2\langle ij\rangle^2$. These two factors, together
with the common helicity numerator $\langle1a\rangle^6$, fix the sign
and normalization of~\eqref{cut:doubleleaf}.

Write $L=\log(s/\mu^2)$, $L_s=L-i\pi$, $L_x=L+\log x$ and
$L_{1-x}=L+\log(1-x)$. The six ordered massless boxes at four points give
\begin{equation}
 \frac{F_4^{(1)}}{T_4}=2sK(x),\qquad
 K(x)=L_xL_{1-x}-x L_sL_{1-x}-(1-x)L_sL_x .
 \label{cut:kernel}
\end{equation}
For channels $S,T$ the finite massless box is $-L_S L_T$ up to a
common constant, with combined coefficient $-2U$ for the two orderings.
The constants drop out by $S+T+U=0$. Inserting~\eqref{cut:azimuth}
and using $K(1-x)=K(x)$ gives an exact integral for the entire moment,
\begin{equation}
 \mathcal C_4=4s^2\sum_j\left[
   \int_{1/2}^{\theta_j}\frac{A_j(x)K(x)}{x(1-x)}\,dx
  -\int_{1/2}^1\frac{(A_j(x)-A_j(1))K(x)}{x(1-x)}\,dx\right].
 \label{cut:full}
\end{equation}
The partial fractions in~\eqref{cut:doubleleaf} reduce the primitive to
iterated integrals of length at most three in the letters
$0,1,\zeta_{ji}$. Equation~\eqref{cut:full} also specifies a direct
integration at finite opening.

\subsection{The double logarithm on the Feynman continuation}

Continue from the timelike sphere to a regular transverse real split
opening with distinct positive angle nodes, $b_j(0)=-c_j(0)\ne0$,
and $s=\kappa t+O(t^2)$, $\kappa>0$. Define
\begin{equation}
 R_j=[t^{q-1}]A_j(1),\qquad
 \chi_j=-\operatorname{sgn}b_j(0),\qquad S_c=\sum_{j\in P_+}\chi_j R_j .
 \label{cut:signed}
\end{equation}
At a generic timelike reference point the first integral in~\eqref{cut:full}
runs along the oriented real segment from $1/2$ to $\theta_j\in(0,1)$,
with $\log x$ and $\log(1-x)$ real there. Continue the contour with its
endpoint, deforming it around the moving poles $\zeta_{ji}$. At the real
split boundary we fix its winding by the contour used in the quadrature.
For $d_j=|\theta_j-1/2|$, follow $x=1/2+i\chi_j r$ from $r=0$ to $d_j$,
then the quarter-circle of radius $d_j$ centred at $1/2$ to $\theta_j$
in the same half-plane, with no extra turns around the real poles.
This gives $\log\theta_j=L_{b_j}-L_{U_j}$ and
$\log(1-\theta_j)=L_{c_j}-L_{U_j}$, where $L_d=\log((-d-i0)/\mu^2)$.
If $U_j$ vanishes identically, take the infinite-ray limit. The second
integral in~\eqref{cut:full} stays on $[1/2,1]$.

\begin{theorem}[The four-point subloop on the negative-pair cut]
\label{cut:theorem}
For $n\ge6$ under these conditions,
\begin{equation}
 \mathcal C_4 =
   4i\pi\kappa^2 S_c\,t^{n+4}\log^2 t
   +O\!\left(t^{n+4}|\log t|\right).
 \label{cut:head}
\end{equation}
If $S_c\ne0$ this is the leading term of the endpoint-subtracted moment.
\end{theorem}

For the cut in figure~\ref{fig:two-loop-cuts}(a), the four-point kernel
on the ray $x=1/2+i\chi_j r$, $1\le r\le\delta/t$, contributes
$-i\pi\chi_j L_s\,dx/x$. The factor $L_s=\log t+O(1)$ supplies one
logarithm, and the integral of $dx/x$ out to radius $\delta/t$ supplies
the other. The residues in the two half-planes enter with opposite signs,
giving the sum $S_c$.

\begin{proof}
For each ordered residue $t\zeta_{ji}$ has a finite nonzero limit, by
Schouten's identity and transversality. There are $n-5$ positive
square brackets in the remaining half-soft tree. Thus
$G_{ij}=O(t^2)$, $B_{ij}=O(t^{q-2})$, and
$A_j(1)=t^{q-1}(R_j+O(t))$. For a sufficiently small fixed
$\delta>0$ we have $A_j(x)-A_j(1)=O(t^q|x-1|)$ when $|x|<\delta/t$.
The second integral in~\eqref{cut:full} is then
$O(t^q(1+|\log t|))$.

Take $x=1/2+i\chi_j r$ on the outgoing ray. Set $u=\log x$ and
$v=\log(1-x)$. The rearrangement
$K=(L+u)(v+i\pi)+xL_s(u-v)$ separates the term linear in $x$.
On $1\le r\le\delta/t$, $u-v=i\pi\chi_j+O(r^{-1})$ and hence
\[
 \frac{K}{x(1-x)}
 =-\frac{i\pi\chi_j L_s}{x}
 +O\!\left(\frac{(1+|\log t|+\log r)(1+\log r)}{r^2}\right).
\]
Replacing $A_j(x)$ by $A_j(1)$ costs $O(t^{q-1}|\log t|)$ in
the integral. The surviving $dx/x$ contributes $\log(1/t)+O(1)$
and $L_s=\log t+O(1)$. This fixes both the sign and the factor $i\pi$.

For the remaining arc put $X=tx$, $Z_t=t\zeta_{ji}$,
$X_t=t\theta_j$. Here $|X_t|$ is bounded away from zero.
On a bounded arc away from zero the kernel differential is
$O(1+|\log t|)|dX|$. If $X_t$ approaches $Z_t$, the identity
\[
 G_{ij}=\frac{(b_j-U_j\zeta_{ji})(b_i-U_i(1-\zeta_{ji}))}
                  {\zeta_{ji}(1-\zeta_{ji})}
\]
gives $G_{ij}/t^2=O(|X_t-Z_t|)$. It supplies the same zero in
$B_{ij}/t^{q-2}$, which absorbs the endpoint logarithm and gives the
continuous value at exact coincidence. Beyond all scaled poles
$A_j(X/t)/t^{q-1}=O(1/|X|)$. The leading kernel term
there is $O(1+|\log t|)|dX|/|X|$, and the integrable remainder is
$O(t)(1+|\log t|+\log|X|)^2|dX|/|X|^2$.
These bounds include infinite endpoints and give at most
$O(t^{q-1}|\log t|)$ before the prefactor $4s^2$.
Multiplication by that prefactor proves~\eqref{cut:head}.
\end{proof}

On the same logarithm sheet, one counterclockwise circuit about an
isolated pole $\zeta=\zeta_{ji}$ adds
$8\pi i\,s^2 B_{ij}K(\zeta)/[\zeta(1-\zeta)]$ to $\mathcal C_4$.
Since $t\zeta$ tends to a finite nonzero limit,
$\log\zeta-\log(1-\zeta)=O(1)$ and~\eqref{cut:kernel} gives
$K(\zeta)/[\zeta(1-\zeta)]=O(t|\log t|)$. With $B_{ij}=O(t^{q-2})$,
any fixed number of such turns changes the moment by at most
$O(t^{n+4}|\log t|)$ and preserves the double-logarithmic coefficient.

The two six-point paths give the following values in their original
external spinor normalization. For the numerical column, integrate
\eqref{cut:full} at $t=10^{-14},10^{-22},10^{-30}$ and take the
centered second difference of $t^{-10}\mathcal C_4$ in $\log t$,
divided by twice the squared step.
\begin{center}
\setlength{\tabcolsep}{10pt}
\begin{tabular}{lcc}
\toprule
Path & $4\kappa^2 S_c$ & Relative error \\
\midrule
Polynomial & $-2067062063104/1865173466513671875$ & $2.90\times10^{-13}$ \\
Uniform exponential & $26270702367/7540720$ & $2.03\times10^{-13}$ \\
\bottomrule
\end{tabular}
\end{center}
The exact column follows from the rational residues in~\eqref{cut:signed}.
An independent integration of the unexpanded spanning trees in both
angles at a timelike $s=1$ point gives relative agreement
$7.42\times10^{-5}$ with the azimuthally reduced integral there.
For the uniform family, write the two arc sizes as $m,k_{\rm arc}\ge2$.
Appendix~\ref{app:cut-cert} proves that $S_c\ne0$ for every $m$ and
$2\le k_{\rm arc}\le1024$. Among right arc sizes up to $H$, at most
$O(H^{1/3})$ remain undecided for any $m$. In particular
$k_{\rm arc}=2$ realizes~\eqref{cut:head} with a nonzero coefficient
at every multiplicity.

\subsection{The two endpoint subtractions for soft and virtual poles}
\label{cut:soft}

The angular moment with a four-point tree in~\eqref{cut:periods} can
be integrated in logarithms. Set $W_{ij}=B_{ij}/[\zeta_{ji}(1-\zeta_{ji})]$.
Integrating the same step functions as in~\eqref{cut:azimuth}, with
constant kernel, gives
\begin{align}
 \mathcal B_{\rm sph}=-2s\sum_{i<j} W_{ij}
 \left[\log\frac{b_i b_j}{G_{ij}}+
       \zeta_{ji}(L_{c_j}-L_{b_j})+
       (1-\zeta_{ji})(L_{c_i}-L_{b_i})\right] .
 \label{cut:treesph}
\end{align}
At the reference point the first log is real. The two ordered residues
contribute conjugate log arguments there, whose product is
$b_i b_j/G_{ij}$. At split data this log denotes
$L_{b_i}+L_{b_j}-\log((G_{ij}-i0)/\mu^4)$. The sum rule
$\sum_{i<j}W_{ij}=\mathcal T$ follows from~\eqref{cut:azimuth}.
The first log supplies $-2\log t$ since $G_{ij}=O(t^2)$, with any
higher-order zero suppressed by its factor in $W_{ij}$.
The remaining terms sum to $\sum_j A_j(1)(L_{c_j}-L_{b_j})$ inside
the brackets with weights included. The log difference tends to
$-i\pi\chi_j$. It follows that
\begin{equation}
 \mathcal B_{\rm sph}
 =2i\pi\kappa S_c\,t^q+4\kappa T_0\,t^{q+1}\log t+O(t^{q+1}).
 \label{cut:treehead}
\end{equation}

Let $E$ denote the external list, $L=(p_1,p_a,-Q_1,-Q_2)$ and
$R=(Q_1,Q_2,\{p_j\}_{j\in P_+})$. Write
$\sigma_H=\sum_{i<j\in H}s_{ij}L_{s_{ij}}$ for each conserving list.
Let $E^{[-1]}$ be the soft radial $1/\epsilon$ coefficient in the
three-particle cut of figure~\ref{fig:two-loop-cuts}(b), integrated over
the hard sphere after taking its residue.
We normalize the soft solid angle to unit area and use
$\rho=2(p_1+p_a)\cdot k/s$ for the dimensionless soft energy.
At fixed hard angles its density is
\begin{equation}
 8\frac{dE^{[-1]}}{d\nu}
 = T_4 T'_n(\sigma_E-\sigma_L-\sigma_R-2i\pi s).
 \label{cut:softdensity}
\end{equation}
To obtain it, write $k=\rho k_0$. Sewing the two leading graviton soft
factors~\cite{Weinberg} gives the following coefficient of $\rho^{-2}$,
\[
 -T_4T'_n\sum_{i\in L,\,j\in R}
          \frac{s_{ij}^2}{s_{i k_0}s_{j k_0}} .
\]
In the rest frame a pair at angular separation
$A=1-\mathbf n_i\cdot\mathbf n_j$ contributes $(2/s)s_{ij}$ times
the angular kernel $A/[(1-\mathbf n_i\cdot\mathbf n)
(1-\mathbf n_j\cdot\mathbf n)]$. Adding $\eta$ to each denominator
and Feynman parametrizing gives the average
\[
 A\int_0^1\frac{dz}{\eta(2+\eta)+2Az(1-z)}
 =\log\frac{2A}{\eta(2+\eta)}+o(1).
\]
Conservation cancels the regulator and all single-leg energy logs.
The cross sum $\sum_{L,R}s_{ij}\log|s_{ij}|$ is the external sum
minus the two subamplitude sums, since opposite copies of a port cancel.
Positive pair invariants in each reference list sum to $2s$,
fixing the phase $-2i\pi s$. The radial measure contributes $s/8$
times $\int_0^{\rho_{\max}}\rho^{-1-2\epsilon}d\rho$ at pole order.
Together these factors give~\eqref{cut:softdensity}.

Define two convergent relative poles and the jointly subtracted pole,
\begin{align}
 D_R&=\int d\nu\,T_4 T'_n(\sigma_R-\sigma_E),&
 D_L&=\int d\nu\,T_4 T'_n(\sigma_L+2i\pi s),\nonumber\\
 P^+&=\int d\nu\,T_4
  [T'_n(\sigma_R+\sigma_L)-\mathcal T(\sigma_E-2i\pi s)] .
 \label{cut:relpoles}
\end{align}
Here $\sigma_R-\sigma_E$ vanishes at least as the square root of the
distance to each endpoint. With
$h_x=x\log x+(1-x)\log(1-x)$ one has $\sigma_L=-2s(h_x+i\pi)$.
Equation~\eqref{cut:softdensity} now gives the two exact identities
\begin{equation}
 8E^{[-1]}+D_R=-D_L,\qquad
 8E^{[-1]}+P^+=(\sigma_E-2i\pi s)\mathcal B_{\rm sph}.
 \label{cut:poles}
\end{equation}
There is a further identity for $P_4=\partial_{\log\mu^2}\mathcal C_4$.
Since $\partial_{\log\mu^2}K=-(h_x+i\pi)$ and
$\int_0^1 h_x\,dx/[x(1-x)]=-\pi^2/3$, one finds
\begin{equation}
 P_4=D_L+\frac{2\pi^2}{3}s^2\mathcal T-2i\pi s\mathcal B_{\rm sph}.
 \label{cut:scale}
\end{equation}
Differentiating the kernel on the intermediate ray in~\eqref{cut:head}
gives $i\pi\chi_j\,dx/x+O((1+\log r)/r^2)|dx|$. On the outer arc
the scaled integral stays bounded. Thus
$P_4=-4i\pi\kappa^2 S_c t^{q+1}\log t+O(t^{q+1})$.
On the original polynomial and exponential six-point paths,
\eqref{cut:poles} and~\eqref{cut:scale} give a first term
$4i\pi\kappa^2 S_c t^{10}\log t$ in $8E^{[-1]}+D_R$. On these same
paths the jointly subtracted pole $8E^{[-1]}+P^+$ starts at $t^9$,
with coefficient $2i\pi\kappa\sigma_E(0)S_c$ and
$\sigma_E(0)=-i\pi\sum_j|b_j(0)|$. The joint subtraction retains
the tree period $\mathcal B_{\rm sph}$ in~\eqref{cut:poles}, whose
leading power on these paths is $t^9$ by~\eqref{cut:treehead}.

\section{Discussion}
\label{sec:discussion}

Near a split-signature helicity zero, the integrated one-loop finite part
cancels all powers below the generic tree order $q=n+3$. At order $q$, the
real parts of opposite hard-pair logarithms cancel, leaving a purely
imaginary single logarithm controlled by the tree cofactor $T_0$. When
$T_0$ vanishes, the finite term $B$ at order $q$ can nevertheless survive.
Tracking a simple branch of the tree-zero locus through order $H-1$ then
makes the loop-to-tree ratio $F_1/\mathcal T$ grow as $t^{-H}$, even
though both amplitudes vanish as $t\downarrow0$.

In real $2\to n-2$ Lorentzian scattering, regular MHV tree zeros are
precisely the planar configurations. The integral
representation~\eqref{ir:sign} shows that the one-loop absorptive part has
a definite sign across the entire regular zero locus, so $F_1$ never
vanishes there. Because the tree zero removes the one-loop infrared pole,
the full one-loop amplitude is finite and nonzero, and infrared
exponentiation feeds this finite value directly into the next-to-highest
pole at every higher loop order~\cite{NS}. The seven-point examples of
section~\ref{sec:physical} show how this conclusion changes with the energy
assignment. At seven points with four incoming particles, $F_1$ can
vanish or be nonzero at the same energy signs and excess Hodges corank.

At what power does the complete two-loop finite part start on the
original six-point polynomial and exponential paths at a single
subtraction scale? On each path a change of scale gives a difference
starting at $t^9\log t$. The value at the reference scale drops out
of this comparison. The four-point subloop gives $t^{10}\log^2t$
on the endpoint-subtracted cut. The bound from the paired paths in
section~\ref{sec:overtaking} applies to a different family of momenta.
To determine $p_2=\operatorname{ord}_t F_2$ on the original paths, the
four-dimensional six-particle integrand of Bourjaily et al.~\cite{SixIntegrand}
provides on-shell input. Both subloop placements and the other cuts
must be combined with the dimensionally regulated hard phase space
before extracting the finite part.

\acknowledgments
This work was supported by the National Natural Science Foundation of China
under Grants 12605132 and 62506128. We also acknowledge support from the
Ningxia Natural Science Foundation
(Grant No.~2026AAC030080) and the Research Start-up Fund for Young
Faculty of Ningxia University. Additional support came from the Zhejiang
Provincial Market Supervision Administration (2026) Science and Technology
Program Project (Peak Project) under Grant No.~JF2026004, the Zhejiang
Provincial Natural Science Foundation of China under Grant No.~LQN25F020023,
and the Huzhou Key Research and Development Program under Grant No.~2025YZ23.

\appendix
\section{A nonvanishing test for the uniform exponential paths}
\label{app:cut-cert}

The signed coefficient $S_c$ in~\eqref{cut:signed} can be decided by
a rational sum that involves only the first spinor derivatives. We give
the reduction and the bound on the possible exceptions because uniformity
in the unrestricted left arc length $m$ matters for the all-multiplicity
conclusion.

\subsection{Deleting the minority leaf}

In the positive affine frame $\lambda_i(0)=(x_i,1)$, let $q_i$ be the
first dotted component at the endpoint and $p_i$ the first derivative of the
second dotted component. The uniform path has left parameters
$r_i=(1+i/(m-1))/4$ for $0\le i<m$ and right parameters
$v_l=(1+l/K)/3$ for $0\le l\le K$, $K=k_{\rm arc}-1$. Its nodes are
$x_X(r)=r+1/r$ and $x_Y(v)=1/v-v$. At each node $q_i$ is the preceding
slope minus the following slope on its string. In order from the first
to the last edge, the slope lists are
\[
 \left(1,\left\{\frac1{1-r_i r_{i+1}}\right\},0\right),\qquad
 \left(0,\left\{\frac{v_l v_{l+1}}{1+v_l v_{l+1}}\right\},1\right).
\]
The sole positive $q_i$ is at the last left site $c$, where $x_c=5/2$.
In this frame $b_i(0)=-q_i$. The only nonzero $p_i$ are
$1/3$ at $c$, $1/2$ at the first right site, and $-1/4$ at the last
right site. Since $\sum_i A_i(1)=\mathcal T=O(t^q)$, we have
$\sum_i R_i=0$ and $S_c=2\chi_cR_c$.

Here is a direct test for $R_c$. At $x=1$ the two cut angle spinors
can be taken as $\lambda_1+z\lambda_a$ and $-\lambda_1/z$. Write
$z=z_0+t w$ near their common limiting collinear direction
$z_0=-X_c(0)$. After an affine change of $w$, the first coefficients of
the positive vertex weights are proportional to $w-x_i$ in the affine
spinor frames. This follows by expanding the first cut angle, whose
two components start as $t$ times an affine function of $w$ and a
nonzero constant, respectively. The second cut angle is a unit there.
The $n-3$ positive bracket ratios each supply $t$, the helicity numerator
supplies $t^6$, the total vertex exponent is $-2$, and $dz/z$ supplies
$t$. Thus $[t^{q-1}]A_c(1)$ is proportional by a nonzero factor to the
residue of this tree in $w$ at $x_c$. Only terms in which $c$ is a leaf
contribute. With $e_{ij}=(q_i p_j-p_i q_j)/(x_i-x_j)$, summing its
attachments gives
\[
 \sum_{i\ne c}(x_c-x_i)e_{ci}=q_c\sum_i p_i\ne0.
\]
The remaining Laplacian has edge weights
$(q'_i p'_j-p'_i q'_j)/(x_i-x_j)$, where
$q'_i=(x_c-x_i)q_i$, $p'_i=(x_c-x_i)p_i$ for $i\ne c$.
All remaining vertex denominators are nonzero.

Let $P(u)=\prod_{i\ne c}(u-x_i)$ and write
$A=P\sum_{i\ne c}q'_i/(u-x_i)$, $B'=P\sum_{i\ne c}p'_i/(u-x_i)$.
The two surviving $p'$ values are $-1/12$ at $8/3$ and $-5/12$ at
$5/6$. Thus $B'$ has full degree $\deg P-1$, with roots at its
inactive nodes and one extra root $\xi=85/36$. At each inactive node
$A(x_i)=q'_i P'(x_i)\ne0$. The extra root is distinct from every node
(the right grid test below proves this even for arbitrarily large $K$).
The deleted cofactor is nonzero precisely when $A(\xi)\ne0$.
Indeed if a vector $f$ is null, and $A_f,B'_f$ insert $f_i$ in the
two sums, every residue of $(A B'_f-B' A_f)/P^2$ vanishes by the null
equations. This rational function has no double poles and decays at infinity.
Hence $A B'_f=B'A_f$. Coprimality and the full degree of $B'$ force
$B'_f=dB'$ and $A_f=dA$ for a constant $d$, giving $f_i=d$ at all
nodes. When $A(\xi)=0$, $f_i=1/(\xi-x_i)$ provides a
nonconstant null vector by direct substitution. Finally $\sum_i q_i=0$
gives
\begin{equation}
 \frac{A(\xi)}{P(\xi)}=(x_c-\xi)D_{m,K},\qquad
 D_{m,K}=\sum_{i\in P_+}\frac{q_i}{\xi-x_i}=F_m+Y_K ,
 \label{cut:rationalsum}
\end{equation}
where $F_m$ and $Y_K$ are the left and right sums. This proves the claimed
equivalence $S_c\ne0\ \Longleftrightarrow\ D_{m,K}\ne0$.

\subsection{The cotangent estimate and simultaneous exception count}

Put $g(r)=1/(\xi-r-1/r)$, $r_0=1/4$ and $r_{m-1}=1/2$. Summation
by parts gives
\begin{equation}
 F_m=g(r_0)+\sum_{i=0}^{m-2}
        \frac{g(r_{i+1})-g(r_i)}{1-r_i r_{i+1}},
 \qquad
 F_\infty=g(r_0)+\int_{1/4}^{1/2}\frac{g'(r)}{1-r^2}\,dr .
 \label{cut:leftgrid}
\end{equation}
All factors are smooth on this interval. On a cell of width $h$ with
midpoint $v$, $r_ir_{i+1}=v^2-h^2/4$. Midpoint expansion of both
expressions gives an error $O(h^3)$ per cell and
$F_m-F_\infty=O((m-1)^{-2})$.

On the right put $v(u)=(1+u)/3$, $d(u)=\xi+v(u)-1/v(u)$.
Its unique zero $\beta\in(0,1)$ satisfies $12\beta^2+109\beta-11=0$.
It is irrational and every finite grid avoids it. The charge at an
interior site $u=l/K$ is the difference of two adjacent slopes, namely
$-w(u,1/K)/K$ with
\[
 w(u,\eta)=\frac{2v(u)/3}
 {(1+v(u)^2)^2-v(u)^2\eta^2/9},\qquad
 a_\eta=\frac{w(\beta,\eta)}{d'(\beta)},\quad a=a_0>0 .
\]
The remainder after subtracting $a_\eta/(u-\beta)$ from $w(u,\eta)/d(u)$
has uniformly bounded first derivative on $[0,1]$, and differs from
its $\eta=0$ value by $O(\eta^2)$. Its mesh average equals its integral
at $\eta=0$ up to $O(K^{-1})$. Both endpoint contributions have limits
with this accuracy. For the remaining pole, recurrence and reflection
of the logarithmic derivative $\psi$ of the gamma function give
\begin{align*}
 \sum_{l=1}^{K-1}\frac1{l-K\beta}
 &=\psi(K(1-\beta))-\psi(K\beta)-\pi\cot(\pi K\beta)\\
 &=\log((1-\beta)/\beta)-\pi\cot(\pi K\beta)+O(K^{-1}).
\end{align*}
If $b$ is nearest to $K\beta$, the nonzero integer
$12b^2+109bK-11K^2$ has absolute value at least one. Factoring it
over the two conjugate quadratic roots gives $\|K\beta\|\ge c_0/K$
for a fixed $c_0>0$. Thus $a_{1/K}-a=O(K^{-2})$ changes the
cotangent term by only $O(K^{-1})$. For a constant $C_Y$ independent
of both arc lengths, this gives
\begin{equation}
 D_{m,K}=F_m+C_Y+\pi a\cot(\pi K\beta)+O(K^{-1}) .
 \label{cut:cot}
\end{equation}
Let $b_m\in(0,1)$ solve $\pi a\cot(\pi b_m)=-F_m-C_Y$, and define
$b_\infty$ using $F_\infty$. The inverse cotangent here has bounded
derivative, so $|b_m-b_\infty|=O((m-1)^{-2})$.
For $H\le K<2H$, a zero in~\eqref{cut:cot} requires
$|\{K\beta\}-b_m|=O(1/H)$. Distinct phases $\{K\beta\}$ have circular
separation at least $c_0/H$. The first $M$ centers therefore admit
$O(M)$ candidates. All later centers lie in an interval of length
$O(M^{-2})$ containing at most $O(1+H/M^2)$ further candidates
after enlargement by $O(1/H)$. Take $M=\lceil H^{1/3}\rceil$ and sum over
dyadic intervals. The resulting $O(H^{1/3})$ bound covers every right
arc size up to $H$ for which a zero is possible at \emph{any} $m\ge2$.

\subsection{A finite certificate with unbounded left arc size}

An explicit bound suffices to decide all $2\le k_{\rm arc}\le1024$.
The Lipschitz constant of $(1-r^2)^{-1}$ on $[1/4,1/2]$ is $16/9$,
and $(1-r_i r_{i+1})^{-1}$ lies between its values at the two ends of
a cell. Since $g$ decreases,~\eqref{cut:leftgrid} gives
\[
 |F_m-F_\infty|\le\frac{C}{m-1},\qquad
 C=\frac49[g(1/4)-g(1/2)]=\frac{252}{85}.
\]
For fixed $K$, test $D_{m,K}$ starting at $m=2$. A strict enclosure away
from zero certifies that $m$. If additionally $|D_{m,K}|>2C/(m-1)$,
the bound certifies the same sign at every later $m$. Applying this
test with outward-rounded balls gives a largest stopping index of seven.
For an equivalent calculation with only integer arithmetic, use $Q=10^{24}$.
For each rational summand $z_i=q_i/(\xi-x_i)$ in~\eqref{cut:rationalsum}
compute $\lfloor Qz_i\rfloor$ and $\lceil Qz_i\rceil$ by integer division.
Sum these bounds to get $[d_-,d_+]$ for $QD_{m,K}$. Check
$d_->0$ or $d_+<0$, and stop the left loop when
$d_->2QC/(m-1)$ or $d_+<-2QC/(m-1)$. This independent enclosure gives
\[
\begin{array}{c|rrrrr}
 \text{first stopping }m & 2&3&4&6&7\\ \hline
 \text{number of right arcs}&1006&14&1&1&1 .
\end{array}
\]
The total is $1023$ right arcs. Every left length is covered for each.
Whether any longer right arc admits a zero on this particular family is open.

\section{Planar tree zeros and the sign of the absorptive part}
\label{app:lorentz}

Choose a frame with the two incoming directions at
\(0,\infty\) on the celestial sphere. The outgoing labels \(i\in O\)
have momenta \(q_i(1,z_i)^{\mathsf T}(1,\bar z_i)\), \(q_i>0\), with
distinct finite \(z_i\ne0\). The two beam matrices have a single
diagonal entry each, respectively
\[
 q_0=-\sum_{i\in O}q_i,\qquad q_\infty=-\sum_{i\in O}q_i|z_i|^2,
 \qquad \sum_{i\in O}q_i z_i=0 .
\]
A sum of two or more outgoing vectors is timelike. For beam 0 and a
nonempty proper outgoing subset \(S\), the invariant is
\(-(\sum_{i\notin S}q_i)\sum_{i\in S}q_i|z_i|^2
-|\sum_{i\in S}q_i z_i|^2<0\). Complements exhaust the other channels.

\paragraph{Planarity from the tree.}
Let \(D(w)=\prod_{i\in O}(w-z_i)\). The interpolating momentum
polynomials in the beam frame have affine representatives
\[
 U(w)=D(w)\sum_i\frac{q_i z_i}{w-z_i},\qquad
 V(w)=wD(w)\sum_i\frac{q_i\bar z_i}{w-z_i}.
\]
Their binary degree is \(N-1\). After nonsingular spin-frame and
Vandermonde factors, the reduced Hodges determinant is their
resultant~\cite{Hodges,Cachazo}. This follows by evaluating
\((U(z)V(w)-U(w)V(z))/(z-w)\) on interpolation nodes and clearing
their Vandermonde factors. A zero of the tree is therefore a common root. Because both \(U(0)\)
and \(U(z_i)\) are nonzero and \(V\) has a nonzero leading coefficient, any common root \(w\) must be finite, nonzero, and off the nodes. Rotate \(w,z_i\) by the same phase to obtain
\(\omega_0\in\mathbb R\) and \(\zeta_i\). The equation for \(U\) gives
\(\omega_0\sum q_i\zeta_i/|\omega_0-\zeta_i|^2
=\sum q_i|\zeta_i|^2/|\omega_0-\zeta_i|^2\), whence
\[
 \operatorname{Re}\sum_i\frac{q_i\bar\zeta_i}{\omega_0-\zeta_i}
 =2\sum_i\frac{q_i(\operatorname{Im}\zeta_i)^2}
                    {|\omega_0-\zeta_i|^2}.
\]
The left side vanishes at a root of \(V\). Strict positivity forces
all \(\zeta_i\) real, which is planarity with the two beams.
Conversely, for planar nodes a rotation gives \(x_i\in\mathbb R\) and hence \(V=wU\).
The degree of \(U\) is \(d=N-2\) since \(\sum q_i x_i=0\) and
\(\sum q_i x_i^2>0\). For \(N\ge3\) there is a common root, and the
degree of the common divisor is exactly \(d\).

\paragraph{Sum of the absorptive logarithms.}
At such a planar point choose rational spinors \(\lambda=(1,x)\) for
the finite nodes and \((0,1)\) at infinity, with
\(\widetilde\lambda_j=q_j\lambda_j\). Put
\begin{gather*}
 y_i=q_i x_i,\qquad
 D_S=\sum_{i\in S}y_i,\qquad A_S=\sum_{i\in S}q_i,\qquad
 B_S=\sum_{i\in S}q_i x_i^2,\\
 G(X)=\prod_{i\in O}(X-x_i),\qquad
 K=\frac{\prod_{j\in O\cup\{0,\infty\}}q_j}{\prod_{i\in O}x_i}.
\end{gather*}
The two-beam component has \(\langle 0,\infty\rangle=1\).
With massless corners \(b,c\), set \(w_i=\langle bi\rangle\langle ic\rangle\).
Applying the weighted Cayley tree identity separately on \(M,M'\)
in~\eqref{one:boxes} gives
\[
 C_{bc;M,M'}=(-1)^{|M'|-2}
     \frac{\prod_jq_j}{\prod_{i\ne b,c}w_i}
     \Big(\sum_{i\in M}q_i w_i\Big)^d .
\]
Indeed, \([ij]/\langle ij\rangle=q_iq_j\) and the two corner sums
are opposite. Their two missing powers are supplied by the Gram
determinant \(q_bq_c(\sum_{i\in M}q_i w_i)^2\).

At \(\mathcal T=0\) the channel identity~\eqref{one:residues} cancels
the squared logarithms. In a box the imaginary part of the remaining
kernel, divided by \(\pi\), is the sum of
\(\sigma_X\operatorname{sgn}(u)\log(1-uX)\) over positions with
\(uX<0\). Here \(u=-\langle bc\rangle^2/(\sum_{i\in M}q_iw_i)^2<0\).
On the two-beam configuration a timelike physical channel has a side
consisting only of outgoing legs. For this side
\(1-uX=U_{bb,S}U_{cc,S}/U_{bc,S}^{\,2}\), where
\(U_{bc,S}=\sum_{i\in S}q_i\langle bi\rangle\langle ic\rangle\).
For two outgoing massless corners the timelike indicators are equal
on all four positions whenever the beams share one massive side, so
the product identity for \(1-uX\) cancels their logs (a singleton
entry has unit argument). If the beams are separated, all nonzero
entries are spacelike. With just one beam among the massless corners
the two timelike positions differ by adjoining the outgoing massless
corner. In the factorized logs, \(U_{jj,S}\) and \(U_{bj,S}\) have
the same values on \(S\) and \(S\cup\{j\}\) since \(\langle jj\rangle=0\).
Thus only differences of
\(\log B_S\) or \(\log A_S\) remain in these boxes. Two-beam boxes
also contribute \(\log|D_S|\).

For example the net coefficient of \(\log B_S\) from all corners is
\[
 (-1)^{N-|S|}K\!
 \left[D_S^d-\sum_{j\in O}
     \frac{x_j(x_jD_S-B_S)^d}{G'(x_j)}\right]=0 .
\]
The first term is from the two-beam box. In a beam-0 box with outgoing
corner \(j\) the two possible positions are \(S\) and \(S\setminus\{j\}\).
Their signs compensate the change in the corner size. The sum extracts
the coefficient of \(X^{N-1}\) from \(X(XD_S-B_S)^d\).
For \(\log A_S\) the corresponding interpolation is
\(\sum_j(D_S-x_jA_S)^d/(x_jG'(x_j))=-D_S^d/G(0)\).
These identities include the singleton positions by continuity. Each
proper unordered split in a two-beam box then leaves
\(-4C_{0\infty;S,S^c}\log|D_S|\). With
\(f(x)=x^d\log|x|\), \(f(0)=0\), the rational-frame amplitude is
\begin{equation}
 \frac{\operatorname{Im}F_1^{\rm rat}}{\pi}
   =-2K\sum_{S\subseteq O}(-1)^{N-|S|} f(D_S).
 \label{ir:forward}
\end{equation}
Terms at \(D_S=0\) extend continuously, including coincident Gram
thresholds at regular kinematics.

The last sum is an iterated forward difference with signed steps \(y_i\).
Integrate a derivative over each step. The common centre shift is
\(\sum_i y_i/2=0\), so the resulting expression is
\(\prod_i y_i\) times \(f^{(N)}\) paired with the density \(\rho\)
in~\eqref{ir:sign}. Since \(f^{(d)}=d!(\log|x|+H_d)\) with \(H_d\) the
harmonic number, one integration by parts gives
\(-d!\prod_i y_i\int_{\mathbb R}\rho'(u)/u\,du\).
The convolution of at least three nondegenerate intervals is even with
Lipschitz first derivative, so the integrand is integrable at zero. A
convolution with another centred interval preserves monotonic decrease
on the positive half-axis. Normalization and compact support ensure
strict decrease somewhere. Finally
\(K\prod_i y_i=s_{\rm in}\prod_{i\in O}q_i^2\).
Restoring conjugate spinor normalizations multiplies the rational
component by \(s_{\rm in}^2/\prod_{i\in O}q_i^2\), giving~\eqref{ir:sign}.
The timelike pairs are the incoming pair and the outgoing pairs.
Their invariant sums are each \(s_{\rm in}\), which gives the imaginary
part of \(\sigma\) used in section~\ref{sec:physical}.

\section{Computational checks}\label{app:replay}

The ancillary files in \texttt{anc/supplement/} contain the rational series
calculations, finite-parameter evaluations and reference outputs.
Run \texttt{python replay.py} in that directory with the dependencies
listed in \texttt{requirements.txt}. Each entry in \texttt{cases.txt}
can also be run individually. The comparator executes each entry in a
fresh temporary directory and compares its standard output with the
corresponding file in \texttt{reproduction/logs/}.

The checks use different data for different assertions. The low-order
cancellations use rational Taylor coefficients of the complete box sum.
The additional ten-point check evaluates the causal functions at two
finite positive openings. The checks on the two-loop shift extract a
coefficient of the difference of normalized amplitudes from their pole
data. The angular checks use the integrated four-point kernel, and the
right-arc certificate combines interval evaluations with an analytic bound
on the entire remaining left-arc tail. At the five-point Lorentzian
zero we sum all 30 box discontinuities in rational arithmetic, keeping
the logarithms of the primes as independent terms. At the seven-point Lorentzian
common zero the sum vanishes in a cyclotomic number field, kernel by
kernel. Individual reference values and the two angular integration
precisions appear alongside the analytic results.

\end{document}